\documentclass[twocolumn,twocolappendix]{aastex7}
\usepackage[utf8]{inputenc}

\usepackage{natbib,graphicx,amsmath,amsthm,ulem,color,wasysym}
\usepackage{graphicx}	
\usepackage{amsmath}	
\usepackage{amssymb}	
\usepackage{hyperref,float}
\usepackage[T1]{fontenc}

\defcitealias{CG97}{CG97}

\newcommand{\y}{\color{red}}
\newcommand{\w}{\color{blue}}
\newcommand{\z}{\color[rgb]{0.8,0,0.6}}

\newcommand{\be}{\begin{eqnarray}}
\newcommand{\ee}{\end{eqnarray}}
\newcommand{\beqn}{\begin{eqnarray}}
\newcommand{\eeqn}{\end{eqnarray}}
\newcommand{\bi}{\begin{itemize}}
\newcommand{\ei}{\end{itemize}}

\def\refnew#1{(\ref{#1})}

\def\g{\, \rm g}
\def\K{\, \rm K}

\def\s{\, \rm s}
\def\km{\, \rm km}
\def\cm{\, \rm cm}
\def\m{\, \rm m}

\usepackage{ulem}
\newcommand{\x}{\sout}
\newcommand{\q}{\color{orange}}
\usepackage{lineno}
\linenumbers

\begin{document}

\title{Bouncing Grains Keep Protoplanetary Disks Bright}

\author[0000-0002-6794-7480]{Yansong Qian}
\affiliation{Department of Astronomy \& Astrophysics, University of 
Toronto, Toronto, Canada}\email{yansong.qian@mail.utoronto.ca}
\author[0000-0003-0511-0893]{Yanqin Wu}
\affiliation{Department of Astronomy \& Astrophysics, University of 
Toronto, Toronto, Canada}
\email{wu@astro.utoronto.ca}

\begin{abstract} 
Proto-planetary disks display the so-called size-luminosity relation, where their mm-wavelength fluxes scale linearly with their emitting areas.  This suggests that these disks are optically thick in mm-band, an interpretation further supported by their near-black-body spectral indexes. Such characteristics are seen not only among disks in very young star-forming regions like Lupus (1-3 Myrs), but, as we demonstrate here, also among disks in the much older Upper Scorpius region (5-11 Myrs). 
How can disks shine brightly for so long, when grain growth and subsequent radial drift should have quickly depleted their solid reservoir?
Here, we suggest that the ``bouncing barrier'' provides the answer. Even colliding at very low speeds (below $1\cm/\s$), grains already fail to stick to each other but instead bounce off in-elastically. This barrier stalls grain growth at a near-universal size of $\sim 100\mu$m. 
These small grains experience much reduced radial drift, and so are able to keep the disks bright for millions of years. They are also tightly coupled to gas, offering poor prospects for processes like streaming instability or pebble accretion. We speculate briefly on how planetesimals can arise in such a  bath of 100-micron grains.
\end{abstract}

\section{Introduction}

Sizes of dust particles in protoplanetary disks is a much sought-after quantity. The dominant grain sizes\footnote{By this we mean mass-weighted size.}  govern the degree of dust-gas coupling, influence the disk dynamics, and shape the pathways towards planetesimal and planet formation.  While
multi-wavelengths observation have yielded strong indications that the solids in disks have grown in sizes compared to their predecessors in the interstellar medium \citep[e.g.][]{Beckwith1990, Ricci2010},
unambiguous constraints on grain sizes remain limited.  
This places any theory of planetesimal formation on shaky grounds. 

Dust grains can grow by sticking, as long as their encounter speeds are below a threshold velocity. This latter threshold is commonly adopted to be the fragmentation speed, speed above which catastrophic destruction occurs. 
And since grain relative speeds typically rise with sizes, either due to radial drifts \citep{Weidenschilling1977}, or turbulence wafting \citep{Volk1980,Cuzzi2003}, such a threshold speed allows grain sticking to proceed to the size range of $\sim$ decimeters. So the "fragmentation-barrier" is able to deliver "pebbles", particles that are marginally coupled to gas and can undergo vigorous evolutions such as streaming instability \citep{Youdin2005}, or pebble accretion \citep{Ormel2010}.

Interestingly, theoretical calculations and lab experiments have suggested the presence of another barrier, the ``bouncing barrier''. This lies at a much lower velocity than that needed for fragmentation \citep[see][for a review, also see Appendix \ref{sec:bouncing} for a brief summary]{Blum2018}. Grains that collide at velocities between 
this 
and the fragmentation speed
can neither stick nor fragment. Instead, they simply bounce away in-elastically. 
The concept of a ``bouncing barrier'' was introduced into astronomy by \citet{Chokshi1993}, who considered head-on collisions, and later extended by \citet{Dominik1997} to  
oblique collisions. 
Bouncing occurs because the energy required to break the contacts between two grains is significantly lower than that 
needed for fragmentation. So grain growth 
can be throttled at a low speed. The bouncing behavior has also been found in laboratory experiments \citep[e.g.][]{Kelling2014, Hill2015, Kruss2016, Brisset2017}
and numerical simulations \citep[e.g.][]{Wada2011, Seizinger2013, Arakawa2023, Oshiro2025}, though the dynamics is complex. The threshold velocity for bouncing may depend on parameters such as grain porosity, composition, mass-ratio, and impact velocity. And only a small parameter space has so far been explored \citep[as summarized by][]{Guttler2010,Windmark2012a}. This complication may partially explain the slow uptake  of this concept in the proto-planetary disk community.

But the bouncing barrier is too important to ignore. When applied to proto-planetary disks, this barrier limits grain growth to a much smaller size, typically $\sim 100\mu$m \citep{Zsom2010,Windmark2012a,Estrada2016,Stammler2023,Dominik2024,Oshiro2025}. This has a score of observational and theoretical consequences. For instance, such grains scatter strongly in the ALMA wavebands, possibly explaining the strong scattering \citep{Zhu2019} and high polarization \citep[e.g.][]{Kataoka2016} inferred from real disks.  As another example, 
\citet{Dominik2024} argued that 
small grains
negatively impact the efficacy of the streaming instability, a current favorite for forming planetesimals.

In this work, we aim to furnish one further argument for the adoption of the bouncing barrier. We show that the bouncing barrier can be used to explain two key observational properties of protoplanetary disks.

The first of these is the so-called size-luminosity relation (``SLR'' from now on), where the disk millimeter luminosities are observed to scale quadratically with disk sizes. Such a correlation was 
first reported by \citet{Andrews2010,Tripathi2017} using 
SMA observations of Taurus and Ophiuchus disks, 
and then found to hold also in  ALMA observations of discs in the Lupus \citep{Andrews2018, Tazzari2021b, Guerra2025}, Upper Scorpius \citep{Barenfeld2017}, Taurus \citep{Long2019}, Chamaeleon I \citep{Hendler2020}, Orion Nebular Cluster \citep{Otter2021} and Ophiuchus star-forming regions \citep{Dasgupta2025}. The SLR is akin to the ``main-sequence'' for stars, and the simplest interpretation for it -- that the luminosity scales linearly with the emitting surface -- is that the emission is optically thick.

Another key feature is the millimeter spectral index $\alpha$, typically measured between 1 mm and 3 mm.  Surveys in Lupus \citep{Tazzari2021}, Taurus \citep{Pinilla2014, Chung2025} and Orion \citep{Otter2021} measured low spectral indexes for the majority of disks,  $\alpha_{1-3} \sim 2$. This could be attributed to either optical thick radiation in the Rayleigh-Jeans limit, or to the fact that emission is dominated by large grains ($> 1$mm). Given the above interpretation for the SLR, the optically thick option seems favored. In this work, we argue that the bouncing barrier is essential to keep disks optically thick.

\bigskip

In this work, we compile the continuum fluxes, dust disk sizes and spectral indexes for a sample of disks from different star forming regions (\S \ref{sec: sample}). In addition, we provide new measurements for a  number of disks in the Upper Scorpius (hereafter USco) region. This region is  particularly interesting for our study:  with an estimated age of 5–11 Myr \citep{Pecaut2012}, it offers a glimpse into the last phases of disk evolution; 
disks in this region are predominately compact, which we argue place the strongest constraints on dust evolution. Regarding disk sizes in USco, previous works (at $0.88$mm) have 
reached conflicting conclusions: \citet{Barenfeld2017} argued that Upper Scorpius disks follow the same SLR as younger regions, whereas \citet{Hendler2020} suggest a much steeper relation, indicating optically thin emissions. This discrepancy may arise from the limited resolution in the survey. 
To resolve it, we analyze archival ALMA observations at $2.86$mm (2015.1.00819.S, PI: Ricci, Luca) to measure $\alpha_{1-3}$ and to infer the disk sizes (\S \ref{sec: usco revision}).

With these data at hand, we then proceed to investigate how the global properties of dusty disks evolve with time (\S \ref{sec: evo}). We argue that only models with bouncing barrier can explain the observed disks. We discuss previous studies and implications in \S \ref{sec: discussion}, before concluding in  \S \ref{sec: conclusion}.

\section{A Sample of Proto-planetary Disks}
\label{sec: sample}

We will focus on disk measurements from the Lupus and the USco regions. Lupus is young, with ages between 1 to 3 Myrs \citep{Alcala2014}, while USco have ages ranging from 5 to 11 Myrs \citep{Pecaut2012}. We choose these two regions because there exist multi-wavelength studies, and  measurements of their stellar properties. Where available, we will also contrast these with disks 
from the Orion Nebula Cluster (thereafter ONC) and Ophiuchus, both similar in age as Lupus.

The following global properties will be investigated. First,  disk fluxes at $0.88$ and $3$mm. For disks where only $1.3$mm fluxes are available, we convert them to $0.88$mm assuming a spectral index $2$. All fluxes are also normalized to be at a distance of 140pc. Second, radii for the dust disks. We cite the value $R_{68\%}$, the radius enclosing 68\% of the total flux. 
This is typically derived from visibility measurements, by assuming a surface brightness profile that is either a broken power-law  \citep{Hendler2020} or a Gaussian \citep{Otter2021, Guerra2025, Dasgupta2025}. Third, stellar luminosity. While most SLR to-date have been presented using raw fluxes, it is more appropriate to compare disks by their ``scaled'' fluxes.
In passively illuminated disks \citep[see, e.g.][]{Chiang1997}, disk temperatures scale as $T\propto L_*^{0.25}$, where $L_*$ is the central star luminosity. The millimeter-wave dust emission, on the other hand, lies in the Rayleigh-Jeans limit and scales as $L_{\rm mm} \propto T$, or, $L_\mathrm{mm} \propto L_*^{1/4}$. So to reduce scatter introduced by stars of different spectral types, 
one should scale the observed disk fluxes by $(L_*/L_\odot)^{1/4}$. 

\subsection{Disk Bulk Properties and the SLR}

\begin{figure}
    \centering
    \includegraphics[width=0.95\linewidth]{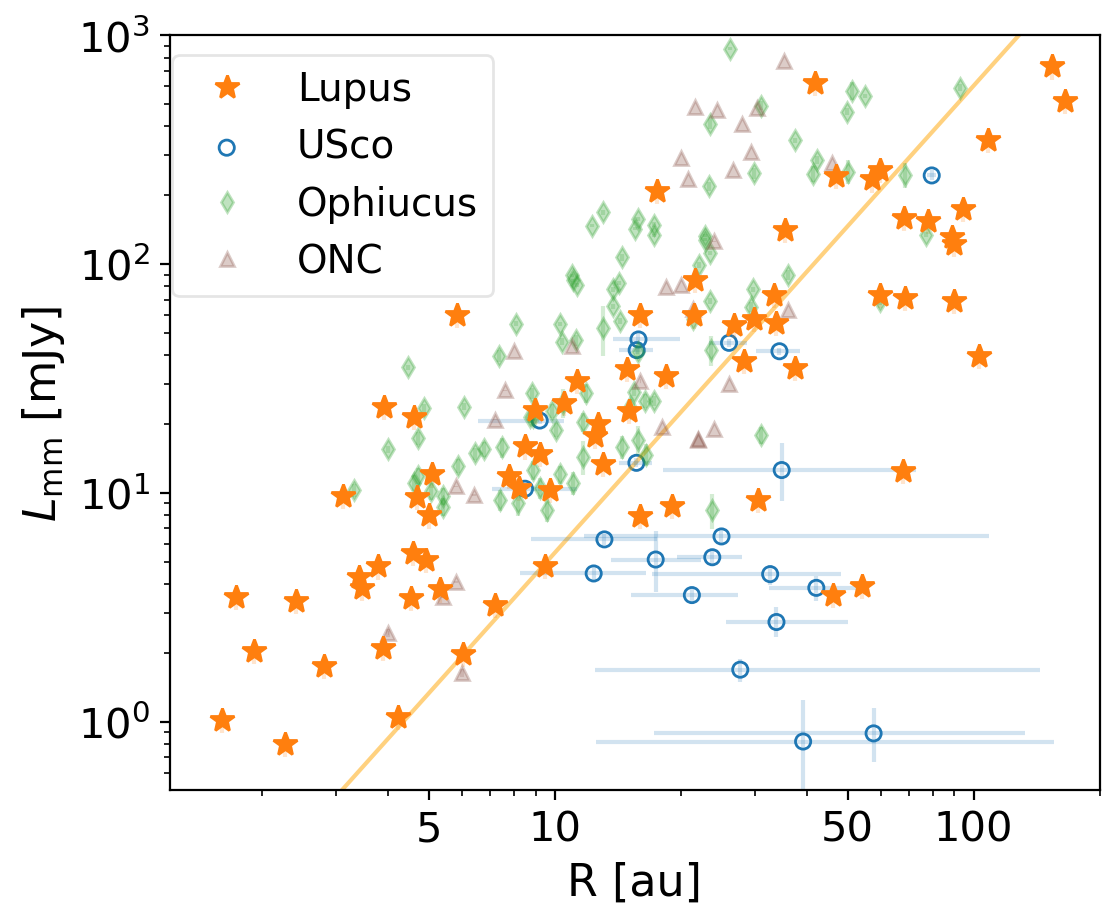}
    \caption{Disk fluxes versus disk sizes, compiled from literature (see text for references).
     The disk radii enclose 68 \% of the total flux. All fluxes are at $0.88$mm and are calibrated to a distance of 140 pc.
  The orange line represents the SLR obtained by \citet{Andrews2018} for large disks in Lupus: $L_{\rm mm}=0.05 {\rm mJy} (R/au)^{2}$. Most disks appear to obey this trend ($L \propto R^2$), except for the USco disks. The Lupus sample (73 disks) have a median resolution of $0.04''$ { ($\sim 6$AU)};
    while the USco sample (21 disks) have a worse median resolution at $0.37''$ ($\sim 58$AU).
    }
    \label{fig:sizel observation}
\end{figure}

\begin{figure}
    \centering
    \includegraphics[width=0.95\linewidth]{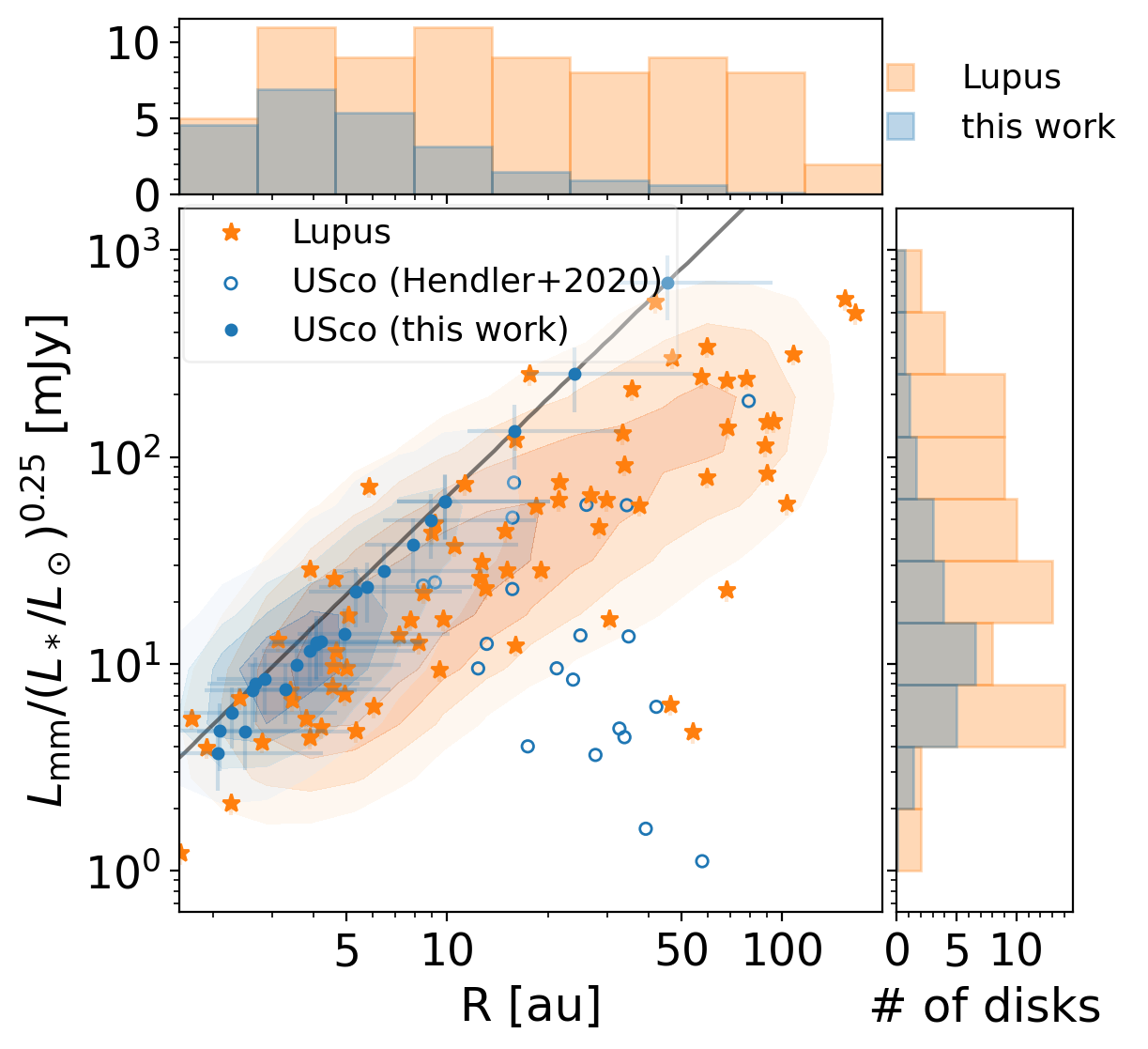}
    \caption{ Same as Fig. \ref{fig:sizel observation} but with disk fluxes scaled by the stellar luminosities.
    Our model radii for 
    USco disks (detailed in text) are shown as solid blue circles. These are typically much smaller than the values from \citet{Hendler2020}.
    The gray line is theory prediction for disks that remain optically thick throughout (at $3$mm)
    and are viewed at an inclination of $60^\circ$ (see Appendix \ref{sec:modelcontinuum}). The Lupus sample suggest that almost all disks are optically thick, with the exception of the largest ones. The same conclusion also holds for the USco disks, if we adopt our model radii. Colored contours are kernel density estimates of disks in Lupus and USco (using our model radii).
    }
    \label{fig:sizel revised}
\end{figure}

\begin{figure}
    \centering   \includegraphics[width=\linewidth]{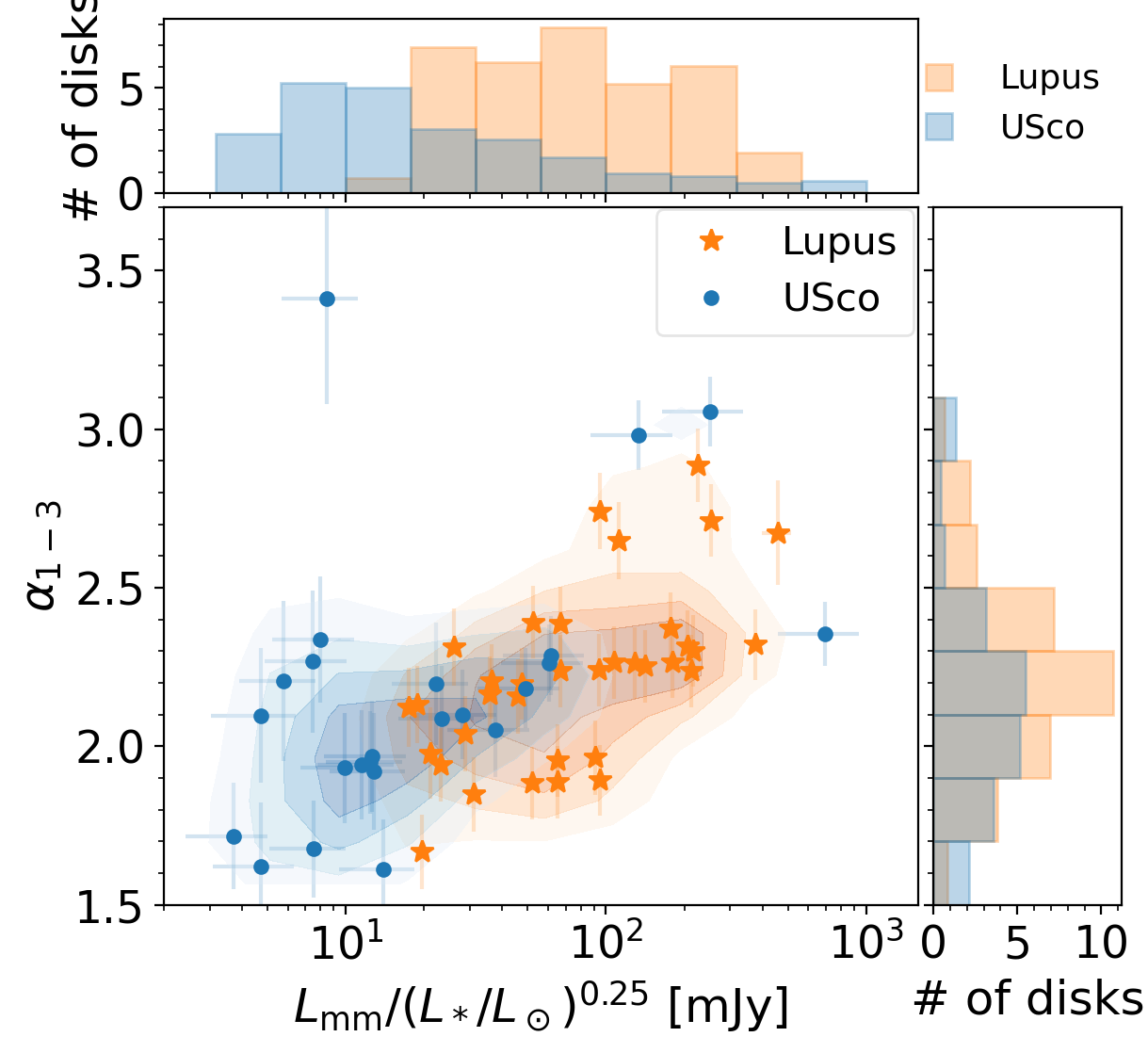}
    \caption{Spectral indexes for disks from Lupus \citep{Tazzari2021} and USco (this work), plotted against the scaled disk fluxes. 
    USco disks are dimmer than those in Lupus  but have similarly low  spectral indexes.
    }
    \label{fig:spi observation}
\end{figure}

We first consider the Lupus region. The high-resolution survey of \citet{Guerra2025} measured fluxes and sizes for 73 disks, in $1.3$mm and at a resolution of $\sim 0.04"$.
Spectral indexes between $0.89$mm and $3.07$mm are provided  for 36 disks in \citet{Tazzari2021}. Stellar parameters are adopted from \citet{Alcala2017} with later correction using Gaia DR2 distance as in \citet{Alcala2019}. Flux measurements are plotted in Fig. \ref{fig:sizel observation} and Fig. \ref{fig:sizel revised} for the unscaled and the scaled fluxes, respectively. A scaling of $L_{\rm mm} \propto R^2$ (the ``SLR'') applies to disks that are as small as a few AU to as large as $\sim 100$AU. Beyond this size, the SLR turns over to a flatter slope.  Spectral indexes of these disks are displayed in Fig. \ref{fig:spi observation}. As mentioned, most disks have low spectral indexes ($\alpha_{1-3} \approx 2$), except for the largest disks. 

We now turn to the USco disks. Surveys in $0.88$mm  \citep{Carpenter2014,Barenfeld2016,Carpenter2024} 
have revealed that these older disks are much less luminous and more compact than those in other star forming regions. Compared to Taurus, for instance,  they are about four times dimmer \citep{Barenfeld2016}, and about three times smaller in sizes \citep{Barenfeld2017}. 
The median spatial resolutions of these surveys are low, $\sim 0.37"$ \citep{Barenfeld2017} and $\sim 0.2"$ \citep{Carpenter2024}. The former corresponds to $\sim 54$AU at a distance of $145$pc. Most detected disks are either un-resolved or just marginally so. This adds complication to the SLR relation. For instance,
while \citet{Barenfeld2017} found that their radius determinations place (at least some) USco disks on the same SLR as younger disks (their Fig. 7), a re-analysis by \citet{Hendler2020} assuming a different surface density profile obtained different radii for many of these disks (Table \ref{tab:spectral}). Plotted against disk fluxes (Fig. \ref{fig:sizel observation}), these radii shift the USco disks away from the said SLR.
We will revisit this issue below. 

We find no published measurements of spectral indexes for USco disks. This leads us to perform our own data reduction on the archival $2.86$mm observations (2015.1.00819.S, PI: Ricci, Luca) of 24 USco disks.
This survey also has a relatively low angular resolution ($2.1''$) but its flux measurements allow us to measure $\alpha_{1-3}$. The publicly available image products from the ALMA archive have been calibrated by the Additional Representative Images for Legacy (ARI-L) Project \citep{ari_l}. 
To measure the disk flux, we apply successively larger apertures until the enclosed flux is constant at a $3\sigma$ level. The uncertainty in the total flux is taken to be the rms of the flux density in the off-field, then integrated over the area enclosed by the aperture. Our measured fluxes are listed in Table \ref{tab:spectral}.
We further adopt stellar parameters for these disks from \citet{Barenfeld2016} and present the scaled fluxes in Fig. \ref{fig:sizel revised}. 

The resulting spectral indexes (between $0.88$ and $2.86$mm) are  presented in 
Table \ref{tab:spectral}. 
These are also plotted in Fig. \ref{fig:spi observation}, showing an average $\alpha_{1-3} \approx 2.2$, with some disks falling below the blackbody value to $\alpha_{1-3} \sim 1.6$.\footnote{This may result from flux uncertainties, or from optically thick disks with high scattering albedos \citep{Zhu2019,Sierra2020}, or from contamination by free-free emission.}

The recent AGE-PRO survey \citet{Agurto2025} returned low-resolution ($\sim 0.3"$) band-6 ($1.3$mm) images of 10 USco disks, 4 of which overlap with our sample. 
Their values  for the spectral indexes (between 1-1.3mm, a narrower range than ours) are largely consistent with our results, except for disk J16082324-1930009 for which they reported $\alpha_{1-1.3} = 3.2$, compared to our value of $\alpha_{1-3}=2.3$.\footnote{We suspect this may be related to their abnormally high 1 mm flux, relative to the literature 0.88 mm and 2.86 mm fluxes.}

Fig. \ref{fig:sizel observation} also displays the ONC \citep{Otter2021} and Ophiuchus \citep{Dasgupta2025} disks for comparison. The spatial resolutions are $\sim 0.03"$  (at $0.85$mm) for ONC and $\sim 0.05"$ for Ophiuchus (at $0.73$mm).\footnote{These corresponds to $\sim 12$AU for ONC (at a distance of $\sim 400$pc), and $\sim 4.5$AU for Ophiuchus (at $\sim 150$pc).} These disks largely span the same ranges in disk sizes and luminosities, and satisfy the same SLR as those in Lupus, though their large disks appear brighter than Lupus disks at the same sizes. This, however, is not a fair comparison. Fluxes would need to be properly scaled by the stellar flux, but we lack stellar parameters in these two regions. In terms of spectral indexes, the average $\alpha_{1-3}$ is $2.1$ \citep{Otter2021} for the ONC and around 2.2 \citep{Tazzari2021} for Ophiuchus, again similar to Lupus and USco.

\subsection{A study of USco disk sizes}
\label{sec: usco revision}

According to the radius measurements from \citet{Hendler2020},  the USco disks do not appear to 
lie on the SLR of other clusters (Figs. \ref{fig:sizel observation}-\ref{fig:sizel revised}). Many USco disks are too dim for their inferred sizes and fall much below the SLR.\footnote{\citet{Hendler2020} instead suggested a much steeper relation of  $L_{\rm mm} \propto R^5$ (their Fig. 6).} This could suggests that these disks are substantially optically thin. Does this spawn from their old ages? or is it a result of measurement uncertainties?

\citet{Hendler2020} obtained  $R_{68\%}$ values by fitting the Band 7 visibilities with a broken-power-law surface brightness profile. Given the low spatial resolution (median $\sim 0.37"$, or 54AU), most of the USco sources show flat visibility profiles, indicating that they are under- or just marginally resolved \citep{Barenfeld2016}. There are margins for error in such a situation.

A stronger argument for measurement uncertainties arises from the spectral indexes we obtain here. As Fig. \ref{fig:spi observation} shows, USco disks follow the same distribution in $\alpha_{1-3}$ as those in Lupus (and ONC). 
Only the brightest  disks show elevated values of $\alpha_{1-3}$, again similar to that in Lupus. This suggests that many USco disks are optically thick,  at variance with their purported departure from the SLR.  We therefore hypothesize that the \citet{Hendler2020} radii are over-estimated.

\begin{figure}
    \centering
    \includegraphics[width=\linewidth]{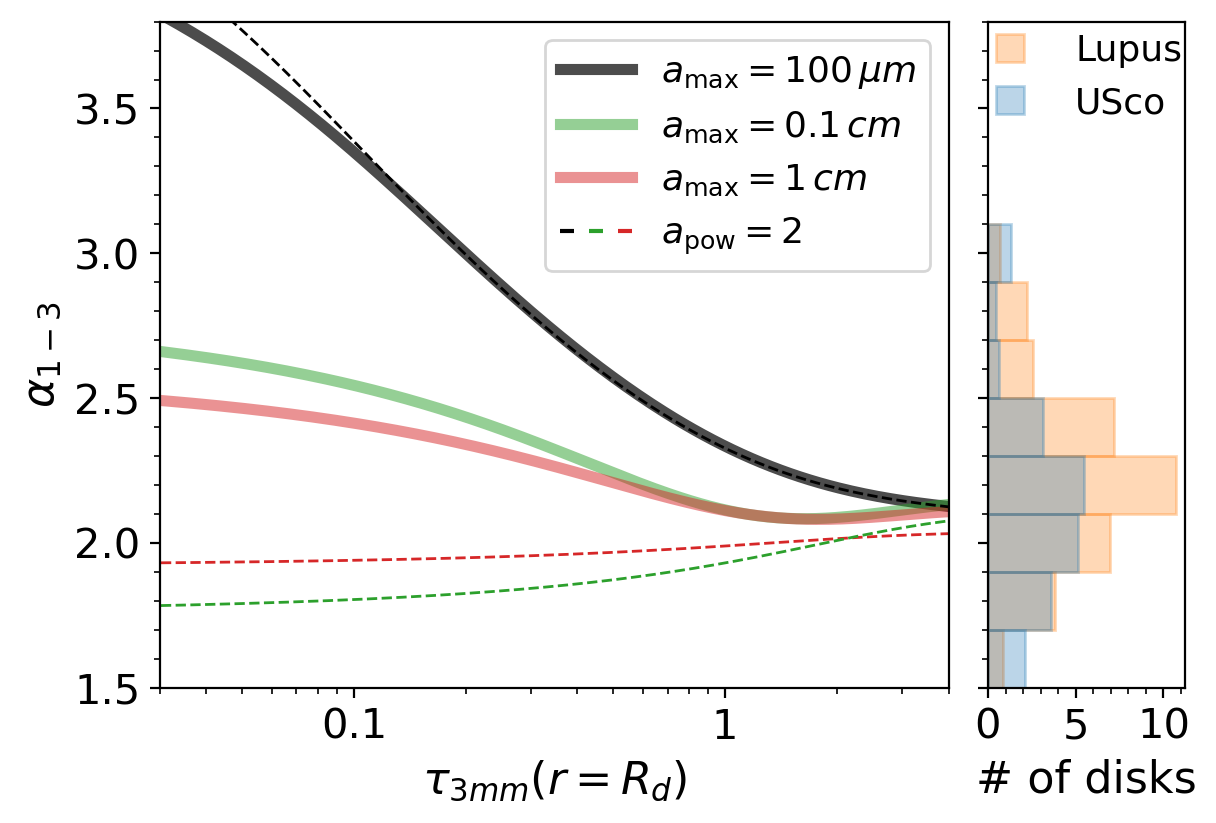}
    \caption{The dependency of spectral index on the optical thickness at the disk outer edge,  $\tau_{\rm 3mm}(r=R_d)$.
    The default dust population (thick black curve) have a maximum size of $a_\mathrm{max}=100 \, \mu m$ and a power-law exponent of $a_{\rm pow}=3.5$.  Other solid curves show different values of $a_{\rm max}$, while the dashed group repeat with $a_{\rm pow} = 2$.  To reproduce the observed distribution of $\alpha_{1-3}$ (right histogram), one either requires optically thick disks (with any $a_{\rm max}$), or optically thin disks with a dust population that is nearly mono-disperse (low $a_{\rm pow}$) and large ($> 1$mm).
    }
    \label{fig:para dependency}
\end{figure}

To proceed, we first perform a quick study of the relationship between spectral index and optical depth.  We will adopt the following truncated power-law profile for the dust surface density, 
\begin{equation}
    \Sigma (r) = \Sigma_{\mathrm{0}} \left( \frac{r}{ 1 \mathrm{au}}\right)^{-p} \exp\left[1-\left(\frac{r}{R_d}\right)^{q}\right] \,,
    \label{eq:density}
\end{equation}
where we restrict ourselves to $p=1$ and $q=5$.\footnote{Our value of $q$ introduces a steeply truncated outer-edge. This helps reproducing the low spectral index, and it is consistent with the fact that $R_{95\%}$ lie close to $R_{68\%}$ in observed disks \citep{Hendler2020, Guerra2025}. }  
We characterize the disk optical depth by that at the outer edge,\footnote{ This is because most of the submm fluxes are contributed by the outer region of a disk. } 
$\tau_{\rm 3mm}(r=R_d) = \kappa_{\rm 3mm} \times \Sigma (r=R_d)$,
and $\kappa_{\rm 3mm}$ includes both scattering and absorption opacities. 
For dust opacities, we adopt the \texttt{DIANA}  form \citep{Woitke2016} and vary both the maximum grain size and the power-law index (details in Appendix \ref{sec:opacity} and Fig. \ref{fig:kappa}).
With that settled, we then use a simple analytical model (Appendix \ref{sec:modelcontinuum}) to compute the integrated disk fluxes, taking into account both dust absorption and scattering. The resulting spectral indexes are presented in Fig. \ref{fig:para dependency}.

From this simple exercise, one finds two possible explanations for a low spectral index.  First, the disk is optically thick throughout ($\tau_{\rm 3mm}(R_d) \geq 1$). Second, the disk can be optically thin, but its thermal emission is dominated by only large grains, i.e., $a_{\rm max} \geq 1$ mm and a top-heavy size distribution (e.g., $\mathrm{apow} \leq 2$). This latter scenario will then be consistent with the large radii deduced by \citet{Hendler2020}. 

However, we suggest the latter explanation may be incorrect. First, the Lupus survey uncovers disks that are similarly dim as the USco disks, yet they satisfy the SLR and are therefore most likely optically thick.
Sizes of these disks are more reliable thanks to a much better spatial resolution ($\sim 0.04"$ as opposed to $\sim 0.3"$ in USco). Second,  it is hard to sustain an optically thin disk that is composed of only large grains -- large grains drain through these disks too fast.\footnote{Sub-structures with local pressure maxima may be invoked to retain these grains. This will be examined in an up-coming contribution.}  Third, a very top-heavy size distribution seems hard to square with the presence of abundant small grains one observes in scattered light. These arguments  lead us to suggest that, the low values of $\alpha_{1-3}$ are emblematic of large optical depths.

If true, then one can infer the physical sizes of the disks based on their measured fluxes alone,\footnote{This assumes that the disks are radially smooth. The referee points out that they could also be made up of many discreet narrow rings. In that case, our results will only represent the ``effective radius''.}
 without spatially resolving them.
Here, with the pre-knowledge of our later sections, we adopt DIANA opacities with $a_{\rm max} = 100\mu$m and $\mathrm{apow}=3.5$. We then use the continuum model in Appendix \ref{sec:modelcontinuum} to search for values of $R_d$ and $\tau_{\rm 3mm}(r=R_d)$ that best reproduce both the observed fluxes in $0.88$mm and $2.86$mm. The disks are assumed to be viewed at $60^\circ$ inclination but we vary the inclination from $33^\circ$ to $81^\circ$ ($1-\sigma$ away from $60^\circ$) to obtain a fiducial 
uncertainty on disk sizes.
From such an exercise, we obtain a new set of 
model 
radii ($R_{68\%}$,Table \ref{tab:spectral},  Fig. \ref{fig:sizel revised}). Since the majority of USco disks have low spectral indexes, our choice of opacity yields large optical depths, typically $\tau_{3mm}(r=R_d) \geq 0.5$. Moreover, the model radii are typically small, much below those reported by \citet{Hendler2020}. 
These disks now lie on the same SLR as defined by the Lupus disks. Such a conclusion, interestingly, seems to be supported by the most recent observations \citep{Pinilla2025}.

We add in a word of caution. Our model radii are not meant to be actual measurements. They are meant to reflect constraints placed by the spectral indexes. 
They are somewhat opacity-dependent and should be replaced when observations with higher spatial resolutions are conducted. In the following section, we adopt these values for our population study.

\begin{table*}
    \centering
    \begin{tabular}{c|c|c|c|ccc}
    2MASS &$L_7$ [mJy]&$L_3$ [mJy]&$\alpha_{1-3}$ & \multicolumn{3}{c}{$R_{68\%}$ [AU]} \\
    & & &  & B17& H20 & model (this work) \\
      \hline
      \hline
J16020757-2257467 & $5.3 \pm 0.3 $ & $0.09 \pm 0.03$ & $3.4 \pm 0.3$ & $32_{- 5 }^{+ 5 }$ & $24_{- 4 }^{+ 4 }$& $3_{ -1 }^{ +3 }$ \\
J16141107-2305362 & $4.77 \pm 0.14 $ & $0.63 \pm 0.08$ & $1.72 \pm 0.17$ & $20_{- 5 }^{+ 6 }$ & $17_{- 3 }^{+ 4 }$& $2_{ -1 }^{ +2 }$ \\
J16181904-2028479 & $4.62 \pm 0.12 $ & $0.47 \pm 0.07$ & $1.93 \pm 0.17$ & $7_{- 2 }^{+ 4 }$ & $12_{- 4 }^{+ 4 }$& $4_{ -1 }^{ +4 }$ \\
J16001844-2230114 & $3.89 \pm 0.15 $ & $0.27 \pm 0.05$ & $2.3 \pm 0.2$ & $20_{- 6 }^{+ 3 }$ & $<12$& $3_{ -1 }^{ +3 }$ \\
J16014086-2258103 & $3.45 \pm 0.14 $ & $0.26 \pm 0.06$ & $2.2 \pm 0.3$ & $24_{- 6 }^{+ 6 }$ & $<14$& $2_{ -1 }^{ +2 }$ \\
J16123916-1859284 & $6.0 \pm 0.3 $ & $0.38 \pm 0.06$ & $2.3 \pm 0.2$ & $33_{- 5 }^{+ 5 }$ & $-$& $3_{ -1 }^{ +3 }$ \\
J16062196-1928445 & $4.1 \pm 0.5 $ & $0.60 \pm 0.07$ & $1.62 \pm 0.20$ & $31_{- 11 }^{+ 22 }$ & $-$& $2_{ -1 }^{ +3 }$ \\
J15582981-2310077 & $5.86 \pm 0.18 $ & $0.59 \pm 0.07$ & $1.95 \pm 0.16$ & $9_{- 2 }^{+ 7 }$ & $25_{- 13 }^{+ 84 }$& $4_{ -1 }^{ +4 }$ \\
J16111330-2019029 & $4.88 \pm 0.16 $ & $0.68 \pm 0.07$ & $1.68 \pm 0.15$ & $5_{- 1 }^{+ 5 }$ & $-$& $3_{ -1 }^{ +3 }$ \\
J16035767-2031055 & $4.3 \pm 0.4 $ & $0.36 \pm 0.06$ & $2.1 \pm 0.2$ & $78_{- 31 }^{+ 60 }$ & $33_{- 16 }^{+ 16 }$& $2_{ -1 }^{ +2 }$ \\
J15530132-2114135 & $5.78 \pm 0.14 $ & $0.59 \pm 0.08$ & $1.94 \pm 0.17$ & $5_{- 1 }^{+ 3 }$ & $13_{- 4 }^{+ 4 }$& $4_{ -1 }^{ +4 }$ \\
J16142029-1906481 & $40.7 \pm 0.2 $ & $3.1 \pm 0.2$ & $2.18 \pm 0.13$ & $20_{- 1 }^{+ 1 }$ & $16_{- 1 }^{+ 1 }$& $9_{ -3 }^{ +9 }$ \\
J16154416-1921171 & $23.57 \pm 0.16 $ & $2.0 \pm 0.2$ & $2.10 \pm 0.14$ & $7_{- 1 }^{+ 1 }$ & $9_{- 3 }^{+ 1 }$& $6_{ -2 }^{ +7 }$ \\
J16153456-2242421 & $11.75 \pm 0.12 $ & $1.16 \pm 0.17$ & $1.97 \pm 0.17$ & $14_{- 1 }^{+ 1 }$ & $35_{- 16 }^{+ 38 }$& $4_{ -1 }^{ +4 }$ \\
J16075796-2040087 & $23.49 \pm 0.12 $ & $2.1 \pm 0.2$ & $2.05 \pm 0.15$ & $7_{- 1 }^{+ 1 }$ & $16_{- 2 }^{+ 4 }$& $8_{ -2 }^{ +8 }$ \\
J16054540-2023088 & $7.64 \pm 0.15 $ & $0.80 \pm 0.12$ & $1.92 \pm 0.18$ & $13_{- 1 }^{+ 3 }$ & $-$& $4_{ -1 }^{ +4 }$ \\
J16135434-2320342 & $7.53 \pm 0.13 $ & $1.13 \pm 0.13$ & $1.61 \pm 0.16$ & $7_{- 2 }^{+ 2 }$ & $-$& $5_{ -1 }^{ +5 }$ \\
J16072625-2432079 & $13.1 \pm 0.2 $ & $0.99 \pm 0.16$ & $2.20 \pm 0.19$ & $20_{- 1 }^{+ 1 }$ & $16_{- 1 }^{+ 1 }$& $5_{ -1 }^{ +6 }$ \\
J16113134-1838259 & $903.6 \pm 0.8 $ & $56.4 \pm 0.5$ & $2.36 \pm 0.10$ & $-$ & $-$& $45_{ -13 }^{ +48 }$ \\
J15583692-2257153 & $174.9 \pm 0.3 $ & $5.2 \pm 0.2$ & $2.98 \pm 0.11$ & $-$ & $79_{- 0 }^{+ 0 }$& $16_{ -4 }^{ +17 }$ \\
J16082324-1930009 & $43.2 \pm 0.8 $ & $3.0 \pm 0.2$ & $2.26 \pm 0.12$ & $44_{- 3 }^{+ 3 }$ & $34_{- 4 }^{+ 4 }$& $10_{ -3 }^{ +10 }$ \\
J16042165-2130284 & $218.8 \pm 0.8 $ & $6.0 \pm 0.3$ & $3.06 \pm 0.11$ & $-$ & $-$& $24_{ -7 }^{ +31 }$ \\
J16090075-1908526 & $47.3 \pm 0.9 $ & $3.2 \pm 0.2$ & $2.29 \pm 0.12$ & $39_{- 3 }^{+ 3 }$ & $26_{- 3 }^{+ 3 }$& $10_{ -3 }^{ +10 }$ \\
J16024152-2138245 & $10.25 \pm 0.19 $ & $0.88 \pm 0.11$ & $2.09 \pm 0.16$ & $16_{- 2 }^{+ 2 }$ & $8_{- 1 }^{+ 3 }$& $6_{ -2 }^{ +6 }$ \\
    \end{tabular}
    \caption{Band 7 \citep{Barenfeld2016} and Band 3 fluxes (this work), spectral indexes measured in between ($\alpha_{1-3}$, this work), and various disk radius determinations ($R_{68\%}$) for 24 USco disks. An additional $10 \%$ calibration error is added to the uncertainties of $\alpha_{1-3}$, because data are taken from two different surveys. The three set of radii are from \citet{Barenfeld2017} (B17), \citet{Hendler2020} (H20), and our model results based on the spectral indexes (see text). The error-bars on our inferred radii 
    include uncertainties from both the flux measurement and the unknown disk inclination.
B17 only provided the truncation radius. We multiply their values by 0.68 to approximate the $R_{68\%}$. 
Our model radii fall substantially below those from B17 and H20 (also see Fig. \ref{fig:sizel revised}).}
    \label{tab:spectral}
\end{table*}

\subsection{Final Sample and Trends}

In the following section, we will be comparing our theoretical models with the Lupus and USco disks, representing respectively young and old populations. We give a brief summary of these disks here.

Our revised radii for the USco disks, 
inferred from a combination of flux and spectral index measurements, place them on the same SLR as for other younger clusters \citep[Fig. \ref{fig:sizel revised} here, also see][for Taurus and Ophiuchus]{Tazzari2021}. If correct, disks across all ages follow a single SLR that has a normalization and slope broadly expected of optically thick emission (Fig. \ref{fig:sizel revised}). This also explains the  same low spectral indexes for disks across all regions.
An exception is found in the largest disks which tend to have higher spectral indexes and lie below the above SLR, suggesting that their optical depths may start to thin down.

In contrast to the younger populations, USco disks are smaller, hardly any larger than 40 au. This may lead one to infer that disks dissipate outside in and that large disks become small over time. On the other hand, USco also have a lower  disk frequency \citep{Ribas2014}. So it is also possible that large disks simply disappear faster, leaving behind only those systems that are initially compact.

\begin{figure}
    \centering    \includegraphics[width=1\linewidth]{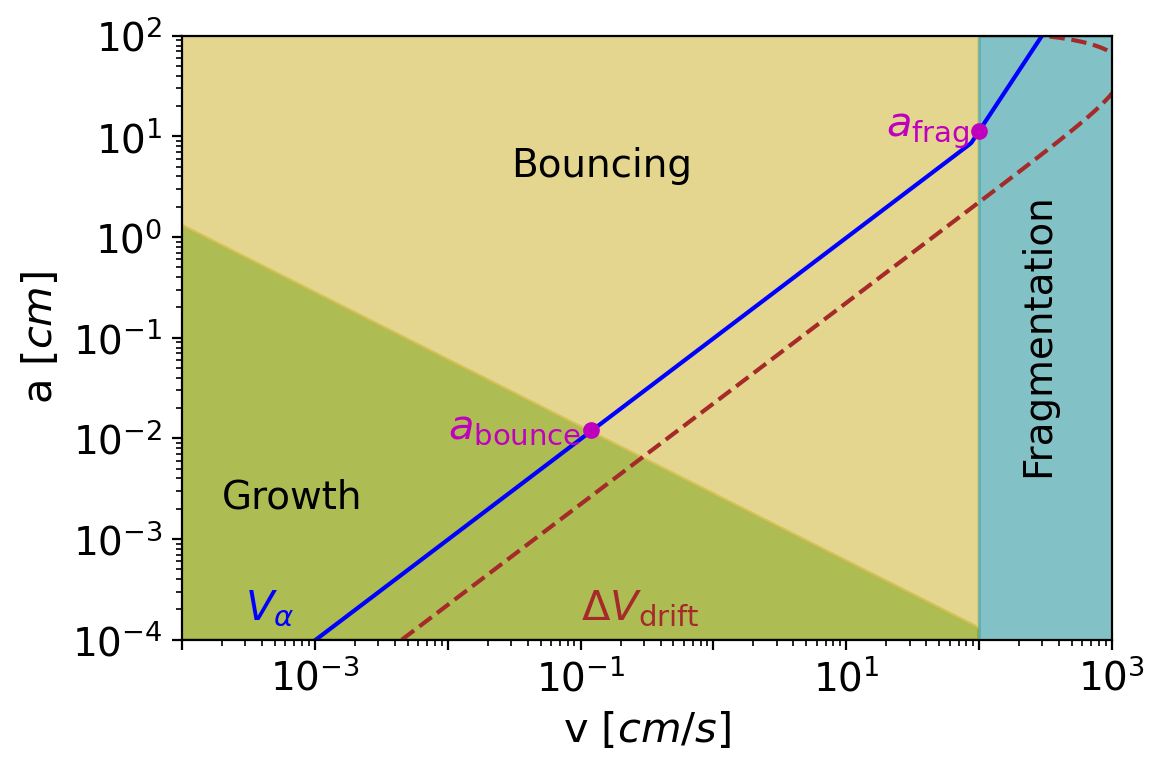}
    \caption{A cartoon illustrating the three regimes of collision outcomes, emulating those in \citet{Guttler2010,Dominik2024}. The regimes are separated by the bouncing and fragmentation thresholds (between two equal-sized grains), as in eqs. \refnew{eq:v bounce} and \refnew{eq:vfrag}. The two lines represent relative encounter velocities, from turbulent stirring, $V_\alpha$ (assuming $\alpha=10^{-4}$, solid curve) and from differential radial drift, $\Delta V_{\rm drift}$ (between grains a factor of 2 in size, dashed curve). These values are evaluated for our default disk at $20$ au. 
    In the presence of bouncing, grain growth is truncated at a small size, $a_{\rm bounce}\sim 100\mu$m, rather than the much larger $a_{\rm frag}$.
    }
    \label{fig:cartoon}
\end{figure}

\section{Modelling the Evolution of Dusty Disks}
\label{sec: evo}

Our above disk sample suggest that, most disks, even the older ones, are optically thick in sub-mm wavelengths. Here, we show that such a fact is difficult to explain in the conventional picture of dust growth and drift. Rather, a mechanism to keep the dust grains small and not drifting is required. The bouncing barrier provides such a mechanism.

In the following, we will first discuss some key ingredients of the model, before presenting results obtained using the \texttt{DustPy} code \citep{dustpy}. 

In this work, we assume that all disks have smooth profiles in density and temperature. This may be problematic. Sub-structures (rings, gaps) in proto-planetary disks are common and exert great impacts on dust migration. However, it is yet unclear if  all disks (especially the smaller ones) host such features. So we choose to focus on smooth disks here. We will turn to the study of sub-structured disks in a subsequent work.

\subsection{Order-of-magnitude Estimates}
\label{subsec:process}

Here, we introduce a few elements of dust evolution that are most relevant to our study.  A more complete picture is presented in, e.g., the recent review by \citet{birnstiel2024}. 

When dust grains collide, the outcome is 
determined by their relative velocity, $\Delta v$.
We consider two most relevant sources to this velocity, turbulent stirring and dust radial drift.

Grain collision speed under turbulent stirring is derived in \citet[e.g.][]{ormel2007} and
summarized in Appendix \ref{sec:dust}. We consider grains with sizes smaller than $a_\eta$, 
where $a_\eta$ (eq. \ref{eq:aeta}) is the size of grains marginally coupled to the smallest eddies in the turbulence cascade. 
Such grains have relative encounter velocities  (eq. \ref{eq:valpha3})
\begin{eqnarray}
V_\alpha & = &  {{\alpha^{3/4} a c_s}\over{\Sigma_{\rm gas}^{3/4}}} \, \left[\rho_{\rm grain} \left({\alpha\over{\beta^2}}\right)^{3/2} \,  \, \left({{\sigma}\over{ 2.3 m_H Re_{\rm min}}}\right)^{1/4}
\right]\nonumber \\
& = & 0.025 \cm/\s \times  r_{\rm au}^{15/28} 
\alpha_{-4}^{3/4}
\left({a\over{100\mu m}}\right) 
\left({{\Sigma_0}\over{2700\g/\cm^2}}\right)^{-3/4}
    \label{eq:valpha}
\end{eqnarray}
where $\alpha$ is the standard Shakura-Sunyaev turbulence parameter \citep{Shakura1973}, with $\alpha_{-4} \equiv \alpha/10^{-4}$. 
We have evaluated the last line of expression using a default disk model of 
$\Sigma_{\rm gas} = \Sigma_0 r_{\rm au}^{-1}$ and $h/r = 0.027 r_{\rm au}^{2/7}$, and a  bulk density for the grains of  $\rho_{\rm grain} = 1.67 \g/\cm^3$ (same below).

In a smooth disk, grains drift inwards radially due to gas drag. The radial drift velocity is 
\citep{Weidenschilling1977}: 
\begin{eqnarray}
    V_{\rm drift}
   &  = & \frac{d\ln P}{d\ln r}
   {{\rho_{\rm grain} a}\over{\Sigma_{\rm gas}}} \left({h\over r}\right)^2\, V_{\rm kep}
   \nonumber \\
   &  = & 0.04  \, \cm/\s\, \times r_{\rm au}^{15/14}\left(\frac{a}{100 \mu m}\right) 
   \left({{\Sigma_0}\over 2700\g/\cm^2}\right)^{-1}\, ,
    \label{eq:vradial}
\end{eqnarray}
where in our default model  $d\ln P/d\ln r= -19/7$.
So depending on the strength of turbulence, the grain's collision speed may be dominated by $V_\alpha$ or $V_{\rm drift}$.

In the absence of a bouncing barrier, grain growth stalls only at the so-called ``fragmentation barrier'', where their relative speed matches the fragmentation limit. We follow previous literature \citep{Blum1993,Guttler2010,Dominik2024} and adopt a constant fragmentation speed of\footnote{This is obtained for mm-sized particles \citep{Blum1993}. It could depend on particle size, as material strength typically decreases with mass.  
\citet{Dominik2024} suggested that it  
could also depend on the size of the monomers in an aggregate.}

\begin{equation}
    V_{\rm frag} = 1 \m/\s\, .
    \label{eq:vfrag}
\end{equation} 
The so-called fragmentation barrier lies at $a = a_{\rm frag}$, which we obtain from setting, e.g., $\Delta V = V_{\alpha} = V_{\rm frag}$,
\begin{equation}
    a_{\rm frag} 
\approx  40  \, \cm \, \times r_{\rm au}^{-15/28} 
\alpha_{-4}^{-3/4} \left(\frac{\Sigma_0}{2700 \,g/cm^2}\right)^{3/4} \, .
    \label{eq:afrag}
\end{equation}
This values scales linearly with $V_{\rm frag}$. So a larger value of $V_{\rm frag}$ \citep[as advocated by, e.g.,][]{Sierra2025} will further increase this size. 

In contrast, the bouncing barrier can throttle grain growth at a much reduced size (see Fig. \ref{fig:cartoon}). We stipulate that grains bounce in-elastically off each other above a speed of (see Appendix \ref{sec:bouncing})
\begin{equation}
    V_{\rm stick}  = 0.2\, \cm/\s \left(\frac{a}{100\mu m}\right)^{-3/2}\, .
\label{eq:v bounce}
\end{equation}
Setting $V_{\alpha} \approx V_{\rm stick}$,
we arrive at (eq. \ref{eq:abounce})
\begin{eqnarray}
a_{\rm bounce}  \approx  220 \mu m \, \times r_{\rm au}^{-3/14} \, 
\alpha_{-4}^{-3/10}
\left({{\Sigma_0}\over{2700 \g/\cm^2}}\right)^{+3/10}\, .
    \label{eq:abounce2}
\end{eqnarray}
This expression depends weakly on almost all disk parameters. So for most disks, it is accurate enough to assign a single value of $100\mu m$ to the bouncing barrier \citep{Birnstiel2018,Dominik2024,Oshiro2025}.  
We can also recast this size in terms of a Stokes number, where  $St = \Omega t_{\rm stop}$ and $t_{\rm stop}$ is the so-called stopping time (eq. \ref{eq:tstop}),
\begin{eqnarray}
    St_{\rm bounce} & = & \Omega_{\rm orb} t_{\rm stop}(a= a_{\rm bounce}) \nonumber \\
    & = & 3.6\times 10^{-5} r_{\rm au}^{11/14}\alpha_{-4}^{3/10}\, \left({{\Sigma_0}\over{2700\g/\cm^2}}\right)^{-7/10} \, .
    \label{eq:stbounce}
\end{eqnarray}
So under the bouncing barrier, grains remain  tightly coupled to the gas.

The above two grain sizes give rise to drastically different disk evolution. We focus on one issue, the timescale for grains to  drift inwards radially.
Grains at the fragmentation barrier are lost rapidly, with
\begin{equation}
    t_{\rm drift,frag} \equiv {r \over{V_{\rm drift}}}
\sim 3000 \mathrm{yrs}\, \times r_{\rm au}^{13/28} 
\alpha_{-4}^{3/4} 
\left(\frac{\Sigma_0}{2700 \,\g /\cm^{2}}\right)^{1/4}\, .
    \label{eq: drift time}
\end{equation}
This timescale is so short, especially for compact disks, it is surprising that we still see such disks at the advanced age of USco region. All solids should have long drained into the star.

In comparison, the bouncing barrier is able to retain the solid much longer, with an estimated drift timescale of
\begin{equation}
    t_{\rm drift,bounce}
    \sim 5.7 \mathrm{Myrs}\, 
    \times r_{\rm au}^{1/7} \alpha_{-4}^{3/10} 
    \left({{\Sigma_0}\over{2700 \g/\cm^2}}\right)^{7/10}\, .
    \label{eq:drift2}
\end{equation}
In this case, even compact proto-planetary disks can remain optically thick for millions of years.

\begin{figure*}
    \centering
\includegraphics[width=0.95\linewidth]{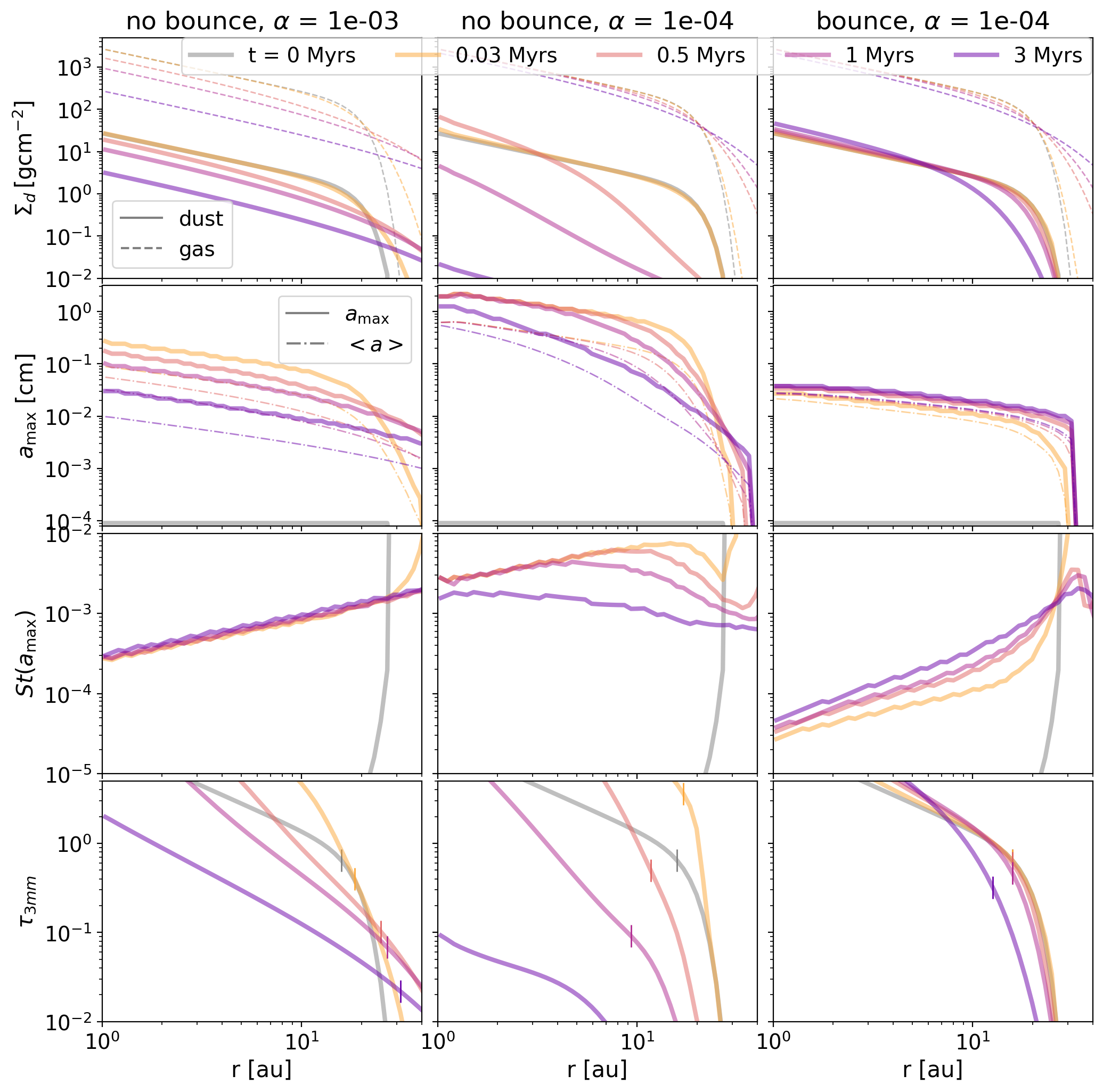}
    \caption{Radial profiles of an evolving disk under different conditions: the two left columns experience different $\alpha$ and no bouncing; the right column is subject to the bouncing barrier.  The disk has an initial density profile  that is cut-off at $R_d = 20$au (eq. \ref{eq:density}). Colored curves show snapshots taken at $t=0,0.03,0.5,1$ and $3$ Myrs. Top row: surface densities of gas (dahsed curves) and dust (solid); second row: 
    maximum (solid) and mass-averaged (dashed) grain sizes, we estimate the maximum one at the top $2.1\%$ ($2-\sigma$) of the mass density; 
     third row: Stokes number for $a_{\rm max}$; bottom row: vertical optical depth ($\tau_{\rm 3mm}$), with the short vertical ticks denoting $R_{68\%}$.  Without bouncing, grains grow and drain rapidly, and the disks turn optically thin within 1 Myrs.
   Bouncing stalls grain growth at $\sim 100\mu$m. 
   This slows down grain drain and the disk remains optically thick and stays at roughly the same size for a few Myrs. 
    }
    \label{fig:sigflux evo}
\end{figure*}

\subsection{DustPy Simulations}

Here we further illustrate the above physics using 1d simulation code \texttt{DustPy} \citep{dustpy}.
This code models the growth of dust grains, as they move under a number of processes: Brownian motion, radial drift, vertical settling, and turbulence. It also models the radial and vertical transport of gas and dust. When two grains collide above the fragmentation speed, they are both disrupted into a power-law spray.
To implement inelastic bouncing into the code, we follow the procedure outlined in Appendix D of \citet{Dominik2024} but with a minor modification as detailed in Appendix \ref{sec:bouncing}.

For our initial disks, we choose the same density profile as in eq. \refnew{eq:density}, with $\Sigma_0=2700\, \g/\cm^2$, $p=1, q=5$, and
initial disk sizes $R_d=[5, 20, 50 , 100]$\,au. 
Our choice of $\Sigma_0$ ensures that even the largest disks are initially optically thick and located on the observed SLR.
The temperature profile follows the \citet{Chiang1997} model of a passively irradiated disk with a host star luminosity of $1L_\odot$ and a constant grazing angle of 0.05 radian. 
The dust to gas ratio is initially 1\% everywhere, and all grains initially  $1\mu$m in size. 
We adopt logarithmic bins in both disk radius and grain mass. The latter spans a range $[10^{-12},10^6]$ g and is covered by 99 grids.
We explore turbulent parameters $\alpha = 10^{-4}$ and  $10^{-3}$. 

\begin{figure}
    \centering
    \includegraphics[width=\linewidth]{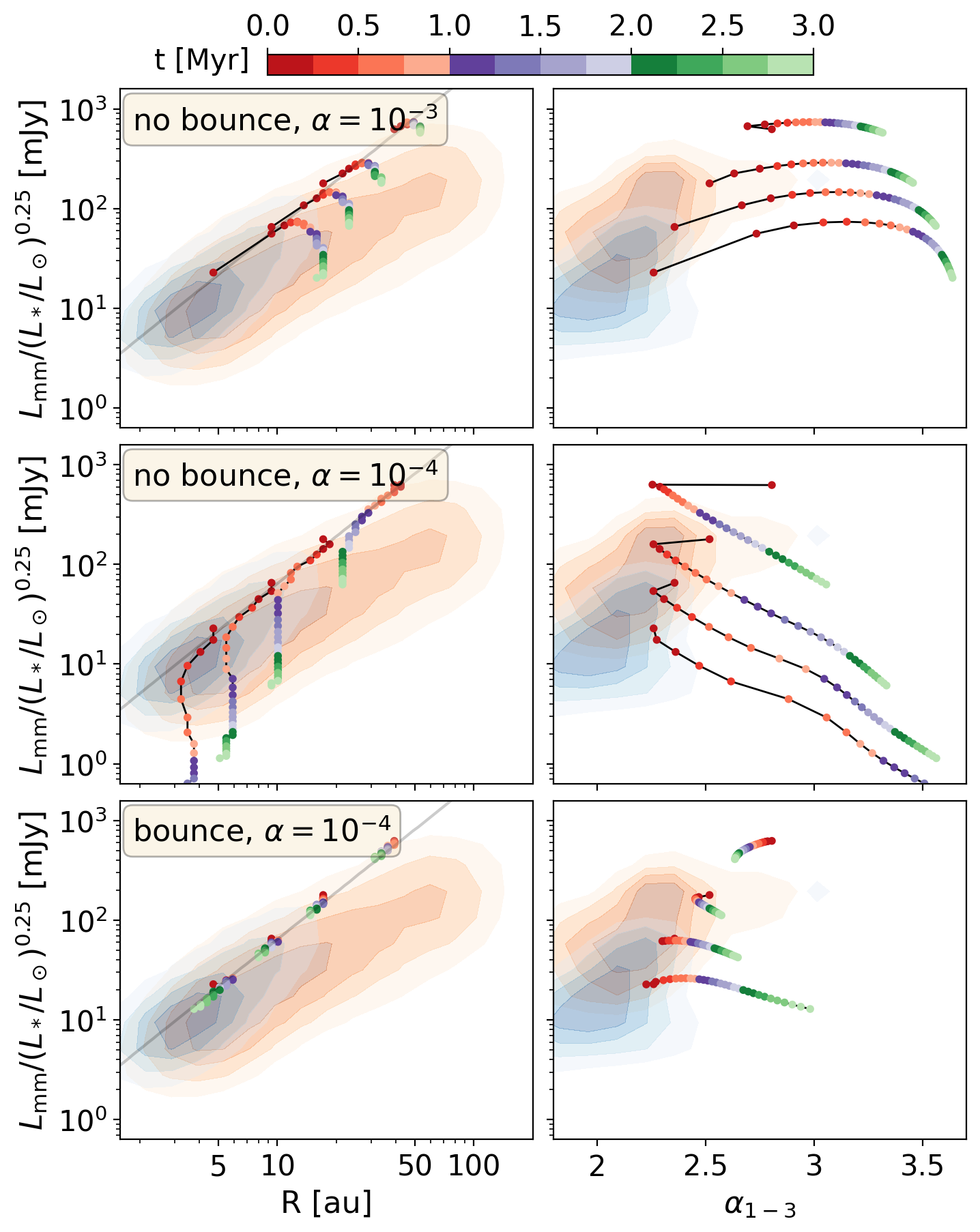}
    \caption{ Evolutionary tracks in the bulk properties (flux, $R_{68\%}$ and $\alpha_{1-3}$) of smooth disks, for the same models as in Fig. \ref{fig:sigflux evo} but with $R_d$ initialized at $5, 10, 20$ and $50$AU. 
    The contours and the black line are as in Figs. \ref{fig:sizel revised} and \ref{fig:spi observation}. 
  Without bouncing, all disks turn optically thin and dim down quickly. It is difficult to maintain these disks near the observed values. 
    Bouncing allows most disks to lie on the observed SLR for Myrs, as well as maintaining small values of $\alpha_{1-3}$. 
The most compact disks do observe a moderate rise in $\alpha_{1-3}$. This results from a reduction in the optical depth by viscous spreading and grain drain. }
    \label{fig:evo}
\end{figure}

The dust opacity is computed using the \texttt{DustPy} output as 
\begin{equation}    \kappa_{tot}=\frac{\int \kappa (a) \Sigma_d(a)\,d\ln a}{\int\Sigma_d(a)\,d\ln a}\,,
\end{equation}
where $\kappa (a)=\{\kappa_\nu^{\rm abs}, \kappa_\nu^{\rm scat}\}$ is the grain opacity for a given $a_\mathrm{max}$, as detailed in Appendix \ref{sec:opacity}. 
To obtain disk fluxes, 
we follow procedure as in Appendix \ref{sec:modelcontinuum} and assume a viewing inclination of $60^\circ$.
Our results are presented in Fig. \ref{fig:sigflux evo} for the radial profiles, and Fig. \ref{fig:evo} for the bulk properties. We describe them in more detail below.

Our choice of dust densities is sufficiently high that dust growth is not limited by the growth timescale,\footnote{ The opposite limit, where dust growth is slow, is invoked by \citet{Powell2017}. However, this requires a low dust density and is incompatible with the bright disks one observes.} but by the fragmentation barrier or the bouncing barrier.

\subsubsection{No bouncing}

In the absence of a bouncing barrier, grain growth proceeds until they reach $a_{\rm frag}$ (eq. \ref{eq:afrag}). Such grains drift inwards quickly (eq. \ref{eq: drift time}). Dust density drops with time dramatically. The decay is more prominent when the level of turbulence is weaker, because the grains can grow to larger sizes. Due to dust drift, after a few Myrs, 
only small grains are retained in the disk's outer part. Disk size (measured as $R_{68\%}$) shrinks with time in the case of weak turbulence. For strong turbulence, it actually increases somewhat as a result of outward advection by the diffusing/expanding gas disk. In any case, all disks become substantially optically thin within 1 Myrs. This has two effects. First, the disk fluxes drop below the SLR, by up to two orders of magnitude. Second, combined with the fact that only small grains persist in the outer parts, the spectral indexes rise to values much above $2$. Both, especially the latter, are in violation of the observed disks.

\subsubsection{Bouncing included}

As expected, the introduction of a bouncing barrier stalls the dust growth at $\sim 100\mu$m (eq. \ref{eq:abounce2}), after just a few hundred orbits.
This dramatically slows down the grain drain. The disks remain optically thick throughout even after a few Myrs. Both disk sizes and spectral indexes remain largely constant.\footnote{Small disks do become marginally optically thin in the outskirts, after a few Myrs. The grain drain there is more severe than we estimated (eq. \ref{eq:drift2}) due to the steeper density profile.} In other words, the bouncing barrier keeps the disks bright.

Among all models, initially compact disks benefit the most from the inclusion of the bouncing barrier. 
This is because the timescale for grain drift is the shortest at small radii (eq. \ref{eq: drift time}). 
Without bouncing, the most compact model ( $R_d=5$ au, Fig. \ref{fig:evo}) experiences rapid evolution in both disk flux and the spectral index.
The fact that not only the young Lupus region exhibit bright compact disks, but also the much older USco region, strongly supports the presence of a bouncing barrier.

\section{Discussions}
\label{sec: discussion}

\subsection{Previous Works on Disk SLR}

We discuss previous works that aim to explain the observed SLR, in the absence of the bouncing barrier.

Following earlier works by, e.g., \citet{Rosotti2019b}, \citet{Zormpas2022} 
presented a large ensemble of 1-D dust evolution models, accounting for many processes including viscosity, grain composition/porosity, fragmentation threshold, gap-opening planet, disk mass etc. They found that their disks often deviate from the SLR while evolving, in agreement with our Fig. \ref{fig:evo}. Compact disks evolve faster and are even harder to explain: their Fig. 6 shows that no disks smaller than $30$au can remain on the SLR for long, consistent with what we found here. They were able to find some particular combinations of parameters that keep the disks on the SLR for a few Myrs,\footnote{This study did not account for disk inclination which will further reduce the disk fluxes.} however, as they show in a later work   
\citep{Delussu2024}, such disks have too large spectral indexes to be compatible with observations. The  spectral index is sensitive to the optical depth in the outer disk, making it an important tracer of disk evolution, in addition to the disk radius. 

That smooth disks should become optically thin quickly is also reported by the study of \citet{Appelgren2025}. 
They further showed that, as a result of the disk turning optically thin, the two flux radii $R_{F95\%}$ and  $R_{F68\%}$, diverge with time, with $R_{F95\%} \gg R_{F68\%}$ at late times
\citep[also see][]{Rosotti2019}. This is contrary to what is found in observed systems \citep{Hendler2020}.

To resolve this tension, 
\citet{Delussu2024} suggested that sub-structures, formed by gap-carving massive planets, are instead responsible for maintaining the SLR and keeping the disk optically thick. This follows previous suggestions that   substructures can effectively stall dust drift \citep[see, e.g.][]{Pinilla2012}. The local pressure maxima forged by these planets (which have to form in a hurry, within $0.1-0.4$ Myrs), they argued, can concentrate and retain large grains. For some parameter ranges, they can reproduce the near 1-1 correlation between $R_{F95\%}$ and $R_{F68\%}$. Most recently, \citet{Kurtovic2025} also argued that sub-structures are necessary to keep the disks bright.

While sub-structures are indeed prevalent in observed disks and are a distinct possibility to explained the SLR, we argue here that the bouncing barrier, a true physical effect, provides a simpler and likely more universal explanation. By stalling grain growth at a small size ($\sim 100\mu$m), the bouncing barrier permits long-lived bright disks, regardless of turbulence strength, disk mass and disk size. To fully support this hypothesis, we will return to consider gapped disks in a subsequent study.

\subsection{The $100\mu$m Grains -- observational evidences}

Our investigation on the SLR and spectral index have led us to believe that grain growth is stalled at around $100\mu$m,  a few orders of magnitude smaller than that expected from the fragmentation barrier. Are there additional observational support for such a scenario? 

The clearest suggestion of such small grains arises from  polarization studies in mm-wavelengths \citep[e.g.][]{Kataoka2016,Yang2016,Hull2018,Ohashi2025}.
These works have argued that the observed polarization is a result of self-scattering or  thermal emission of aligned grains, and that millimeter emission is dominated by grains with sizes $\sim 100\mu$m.

Another line of evidence, though less restricting, is related to the inferred optical depths in observed disks. These appear to plateau at $\tau_{\rm mm} \sim 0.6$. To explain this, \citet{Zhu2019} proposed that these disks are actually optically thick, but grain scattering reduce the apparent optical depths. The data is consistent with a high millimeter albedo ($\geq 0.9$). According to our Fig. \ref{fig:kappa}, this limits the maximum grain size to somewhere in-between $100\mu$m - $1\cm$.
The higher choices ($\sim 1\cm$) may be slightly disfavored, as it would require more dust mass to make the disks optically thick.

In addition, disks that are imaged with sufficiently high spatial resolution and in multiple bands can be used to infer the dominant grain sizes. For instance, \citet{Zagaria2025} found a maximum grain size of $\sim 700\mu$m in the CI Tau disk, while \citet{Yoshida2025}  found a value of $340\mu$m in the TW Hya disk. 

Lastly, one can also infer grain sizes using their vertical distribution and -- in disks with sub-structures -- their radial spread. Conclusions from such studies are less clear-cut, as they are subject to uncertainties in 
disk turbulence and pressure profile \citep[see, e.g.][]{Doi2023,
Jiang2025}.

In conclusion, while many lines of studies support the dominance of small grains, a definitive picture is yet to emerge.

\subsection{Bouncing Barrier -- theoretical issues}

Here, we consider some theoretical issues with regard to the bouncing barrier.

The $100\mu$m size is emblematic of the bouncing barrier. It is robust to changes in the disk environment (eq. \ref{eq:abounce2}).
Disk simulations that incorporated the bouncing barrier \citep{Zsom2010,Windmark2012a,Stammler2023,Dominik2024,Oshiro2025} 
all produced a mono-disperse population of dust grains with $a \approx 100\mu$m,\footnote{\citet{Zsom2010} reported a constant Stokes number of $\sim 10^{-4}$ at 1au. This is compatible with our eq. \refnew{eq:stbounce}.} 
despite adopting different complexities for the collision outcomes, and working under a wide range of disk parameters (radius, turbulence, density...).  

This robustness also extends to the case where one considers a spread of encounter velocities around the mean value. As there are always some collisions that occur at speeds lower than $V_{\rm bounce}$, 
\citet{Windmark2012a} have argued that this could 
allow some lucky grains to break through to much larger sizes. We believe this is unlikely,  As grains grow in size, they are stirred to higher speed by turbulence (or drift), yet they become harder to stick ($V_{\rm bounce} \propto a^{-3/2}$, eq. \ref{eq:v bounce}). This throttles further growth. 

While largely insensitive to the environment, the exact grain sizes are affected by a number of grain properties, e.g., grain porosity, shape, chemical composition and monomer size.
 \citet{Oshiro2025} found that the threshold velocity for (face-on) bouncing scales steeply with the volume filling factor $\phi$ as $v_{\rm bounce} \propto \phi^{-14}$ (their eq. 12). 
This arises because aggregates that are more densely packed have higher compressive strengths. They are thus harder to deform at impact, while easier to tear apart after contact. 
The porosity of real grains are unknown, worse still it may be a function of particle size and collision history.
In addition, if rolling motion determines the bouncing threshold, the rounded-ness of grains also matters. Lastly, the bouncing threshold also depends on the size of the monomers, as well as grain composition. For example, icy grains, with a higher sticking efficiency, can grow to larger sizes  \citep{Chokshi1993}, and possibly with a larger  porosity \citep{Okuzumi2012}.

The robustness of the $100\mu$m grain, ironically, presents a dilemma \citet[also see][]{Dominik2024}.
Simulations find that almost all dust masses are quickly, typically within a few hundred orbits, absorbed into $100\mu$m grains. This leaves little trace of the small grains (micron-sized or smaller), ones that are abundantly observed 
in scattered lights. These grains are seen near the disk surfaces and 
are responsible for absorbing/scattering star light, providing the energy for the big grains' thermal emission. 
Do these small grains somehow avoid integration? or are they resupplied from colliding large grains?\footnote{\citep{Guttler2010} argued that erosion is unlikely at the low-speed of the bouncing collisions.}

\subsection{Life After Bounce?}

Starting from micron-sized grains, within a few hundred orbits, grain growth has come to a halt at the bouncing barrier. What would happen to such a bath of $100\mu$m grains, during the millions of years of disk lifetimes? When do our coveted planetesimals form?

We first consider the streaming instability, the current leading model for planetesimal formation. This instability arises from the relative radial drift between gas and grains.
It grows optimally, when dust grains are marginally coupled to the gas ($St \sim 1$) and when they are highly concentrated towards the midplane by vertical settling \citep{Youdin2005}. 

In contrast, the $100\mu$m grains are strongly coupled to the gas (eq. \ref{eq:stbounce}). With a Stokes number of order $St \sim 10^{-4}$, they experience very small radial drifts. This  reduces the amount of free energy available  to the streaming instability. Both the growth rate and the saturation level of the instability will be much weakened.

Moreover, the instability is hampered by the fact that these small grains are easily wafted up. To make a dense mid-plane layer with comparable mass densities in gas and dust, the grains will have to have an equilibrium scale height $H_p \sim Z\times h$, where $Z \equiv \Sigma_{\rm dust}/\Sigma_{\rm gas}\sim 0.01$. With $H_p$ evaluated as $H_p/h \sim\sqrt{\alpha/St}$ \citep{Dubrulle1995, Youdin2007},
we find that the permissible level of turbulence cannot exceed
\begin{equation}
\alpha \leq  \left(\frac{H_p}{h}\right)^2 \, St \sim 10^{-8} \times \left(\frac{St}{10^{-4}}\right)\, \left(\frac{H_p/h}{10^{-2}}\right)^2\, .\label{eq:alphastir}
\end{equation}
To satisfy this, any fluid instabilities would have to be severely suppressed, including the so-called  vertical-shear-instability \citep[e.g.][]{Nelson2013}. It is unclear if this is possible. This stringent condition leads us to suggest that the streaming instability is unlikely to be effective in the presence of the bouncing barrier.
The same sentiment also extends to other proposed processes like concentration by vortexes or pebble accretion. 

With these proposed pathways blocked, subsequent evolution of the grains is unclear.  
One interesting idea was proposed by 
\citet{Windmark2012a}. They performed a numerical experiment by injecting some cm-sized ``seeds'' into a bath of $100\mu$m grains. They found that,
through mass-transfer collisions,\footnote{These are high speed collisions between grains of large mass-ratios, where the small projectile fragments on impact and implants part of its mass onto the much larger target.} 
these seeds can efficiently sweep up the $100\mu$m grains\footnote{This is analogous to a car-screen swatting flies at high speed.} and grow rapidly.
In fact, they argued that the $100\mu$m bath provides the perfect environment for fueling such growth. 

To illustrate this, we estimate the timescale for a seed of size $a$ to double, assuming perfect accretion. In lieu of gravitational focusing, this is 
\begin{eqnarray}
t_{\rm double} & \approx & {{\rho_{\rm grain} a}\over{\pi \Sigma_{\rm dust}}} \left({h\over r}\right)^{-1} \, P_{\rm orb}\nonumber \\
& \approx & 4.4 {\rm Myrs} \times r_{\rm au}^{31/14} \left({a\over{22\km}}\right) \left({{\Sigma_0}\over{2700\g/\cm^2}}\right)^{-1}\, ,
    \label{eq:double}
\end{eqnarray}
where we have assumed that the seed is moving at the local Keplerian speed, and that the $100\mu$m grains are well coupled to the gas. So they move at a relative speed of $\sim (h/r)^2 v_{\rm kep} \sim 2.2\times 10^3 \cm/s \times r_{\rm au}^{1/14}$. When a seed has grown to a size of $a \sim 22\km$, the relative speed falls below its surface escape velocity. It can then grow exponentially (run-away accretion) with the assist of gravitational focusing. 
In summary, the presence of lucky seeds may give rise to large planetesimals.

\section{Conclusion}
\label{sec: conclusion}

The concept of bouncing has been introduced into the field of astronomy decades ago \citep{Chokshi1993,Dominik1997,Guttler2010}. But despite its importance,
its physical confirmation by experimental and numerical studies, and its observational support by sub-mm polarization, the general uptake of this concept within the broader community appears to be low. 

Inspired by the recent work of \citet{Dominik2024}, we explore how the bouncing barrier is critical to explain the bulk properties of observed disks, i.e., their size-luminosity relation and their low spectral indexes.  We hope our work furnishes yet another argument for adopting the bouncing barrier in the wider community. 

The inclusion of bouncing effectively limits dust growth to a near-universal size of $\sim 100\mu$m. This dramatically slows down the radial drift of dust grains in smooth disks. Without bouncing,  compact (1-10au) disks would have lost all its dust within $\sim 10^4$ yrs (eq. \ref{eq: drift time}), much shorter than the typical lifetimes of disks (1-10 Myrs)
Even for larger disks, their low spectral indexes and their size-luminosity relation indicate that their outer parts remain optically thick in mm-wavelengths, regardless of the ages of the star-forming regions. This is impossible to reach in smooth disks, unless the grains remain small in sizes. 

More observational studies are needed to further confirm the bouncing hypothesis. Theoretically, one particularly interesting direction is to study disks with sub-structures. Given that local pressure maxima may act to prevent dust drift, one needs to evaluate the need for the bouncing barrier in such disks. We plan to do so in our next work.

The fate of bouncing-limited disks is also of high import. Small grains, with their low Stokes numbers, are strongly coupled to the disk gas and are ineffective in triggering the streaming instability. 
So how does a bath of $100\mu$m grains overcome the growth barrier to form planetesimals?

\bigskip
The authors thank Luca Ricci for the ALMA data on Upper Scorpius, and Osmar M. Guerra-Alvarado for sharing Lupus data. We thank  Anders Johansen, Jose Francisco Gomez, Debanjan Sengupta, and an anonymous referee for useful comments on the draft. We also thank Chris Thompson, Chris Matzner, Simin Tong and Xinting Yu for helpful conversations, and NSERC for research support.

This paper makes use of the following ALMA data: ADS/JAO.ALMA\#2015.1.00819. ALMA is a partnership of ESO (representing its member states), NSF (USA) and NINS (Japan), together with NRC (Canada), NSTC and ASIAA (Taiwan), and KASI (Republic of Korea), in cooperation with the Republic of Chile. The Joint ALMA Observatory is operated by ESO, AUI/NRAO and NAOJ. The National Radio Astronomy Observatory is a facility of the National Science Foundation operated under cooperative agreement by Associated Universities, Inc.

\bibliography{ref}

\begin{appendix}

\section{A simple Model for Dust Continuum Emission}
\label{sec:modelcontinuum}

To derive the specific intensity from the disk, we take into account both dust absorption and scattering. Previous studies have shown that scattering not only reduces the total emission but also alters the spectral index due to its strong dependence on wavelength \citep{Zhu2019}. This effect becomes particularly important in optically thick disks.

\citet{Miyake1993} solve the radiative transfer equation for a disk using the two-stream approximation. Here, we follow \citet{Sierra2020} where they further simplify the expression assuming the disk is faced-on and vertically isothermal: 
\begin{equation}
I_{\nu} = B_{\nu}(T_d) \left[ 1 - \exp\left(-\frac{\tau^{\rm abs}_\nu}{1-\omega_\nu^{\rm scat}}\right) + \omega_{\nu} {\cal F} (\tau^{\rm abs}_\nu,\omega_\nu^{\rm scat}) \right],
\label{eq:intensity sc}
\end{equation}
where the absorption optical depth $\tau^{\rm abs}_\nu (r)=\kappa^{\rm abs}_\nu \Sigma(r)$ and the scattering albedo $\omega_\nu=\kappa_\nu^{\rm scat}/(\kappa_\nu^{\rm scat}+\kappa^{\rm abs}_\nu)$. Detailed expression for the factor $F$ is presented in \citet{Sierra2020}.

When the observing wavelength is much longer than the blackbody peak ($\lambda >>\lambda_{peak}$), the blackbody emission can be approximated by the Rayleigh-Jeans law,
\begin{equation}
    B_\nu(T_d)= \frac{2k_bT_d(r) }{\lambda^2} \,,
\end{equation}
where the Boltzman constant $k_b = 1.38\times 10^{-16} \mathrm{erg}/\mathrm{K}$ and $\lambda$ is the observing wavelength. We assume the disk temperature follows a simple power law:
\begin{equation}
    T_d = 150\, \mathrm{K} \left(\frac{L_*}{L_\odot}\right)^{1/4}\left(\frac{r}{1 \mathrm{au}}\right)^{-3/7}\, ,
    \label{eq: temperature}
\end{equation}
where $L_*$ is the luminosity of the host star and $L_\odot$ is the solar luminosity.  

To calculate the continuum flux at the mm-wavelength ($L_{mm}$), 
we integrate the above specific intensity over the disk surface area, 
\begin{eqnarray}
    L_{mm} &=& \int I_\nu d\Omega =  \frac{2\pi}{d^2}\int I_\nu r dr\,,
    \label{eq: L model}
\end{eqnarray}
where adopt a common distance of $d=140$ pc.
To account for the disk inclination, we reduce the above flux by a factor  $\cos\theta$, with $\theta=60^\circ$ for all disks.

\section{Opacity}
\label{sec:opacity}
\begin{figure}
    \centering    \includegraphics[width=\linewidth]{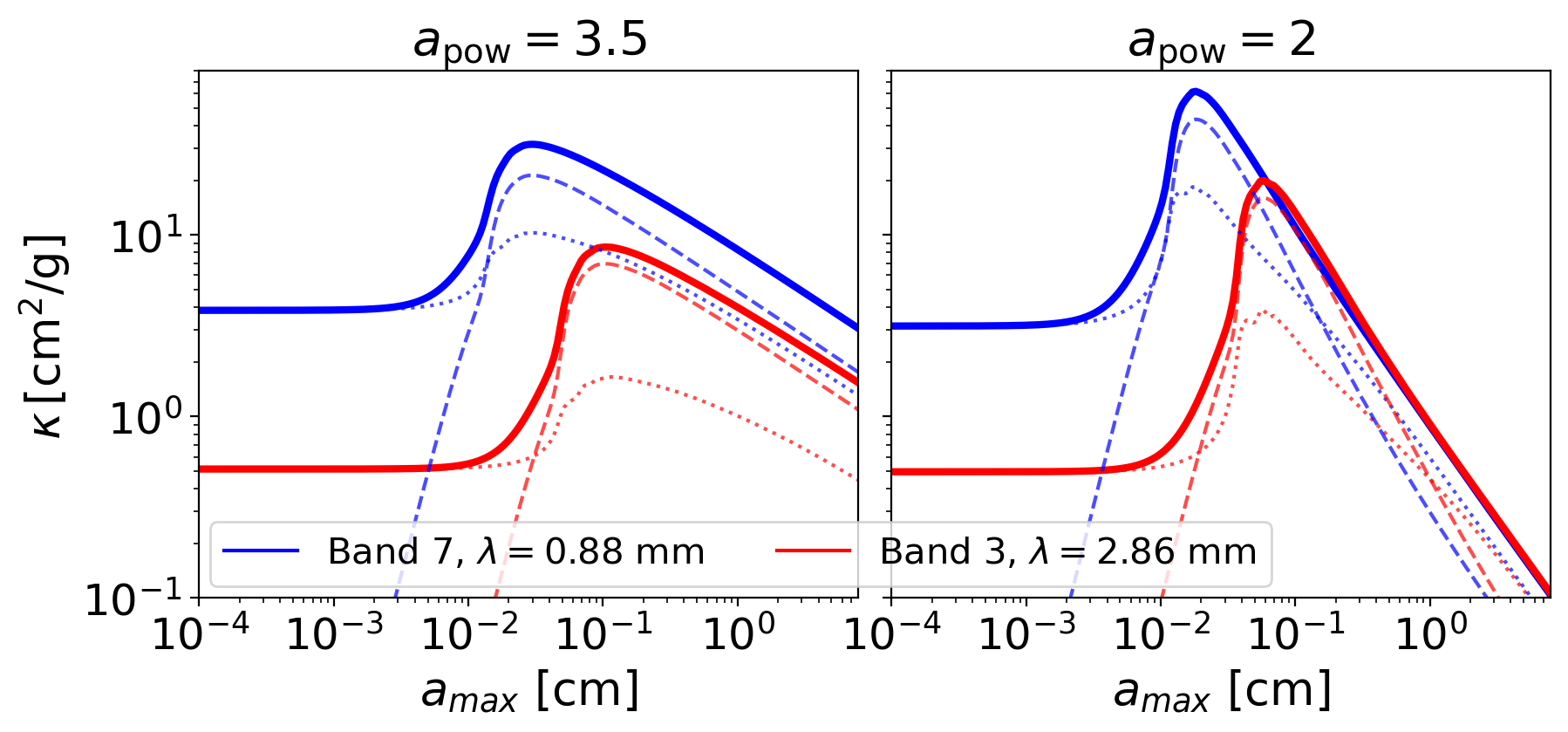}
    \caption{
   Dust opacities (per gram of dust mass) for power-law grain size distributions, $dN/da \propto a^{-3.5}$ at the left and $a^{-2}$ at the right, at two observing wavelengths and plotted against the maximum grain sizes. The solid lines represent the total opacity, while the dotted and dashed that of absorption ($\kappa^{\rm abs}_\lambda$),and scattering ($\kappa^{\rm scat}_\lambda$) components. 
   Scattering dominates for grain sizes of interest ($100\mu$m to cm). The value of $\beta = d\ln \kappa/d\ln\lambda$ depends on the size distribution, being nearly zero for the right panel when $a_{\rm max}$ is beyond $1$mm.  }
    \label{fig:kappa}
\end{figure}

To model the disk luminosity 
we require the absorption coefficient $\kappa_\nu^{\rm abs}$ and scattering coefficient 
 $\kappa_\nu^{\rm scat}$. 
We use the Fortran package \texttt{OpacityTool} to compute the DIANA standard dust opacities \citep{Woitke2016}. These are from grains with a porosity of $25\%$ and a composition of $60\%$ silicates and $15\%$ amorphous carbon (in volume).  We note that DIANA's $\kappa_\nu^{\rm abs}$ is a few times higher than that for Ricci's \citep{Ricci2010} and \texttt{DSHARP} opacities \citep{Birnstiel2018}. However, this discrepancy does not matter in our analysis because the opacity is instead dominated by scattering when the dust size is comparable to the observing wavelength (as is the case here). Throughout the paper, we adopt a grain size distribution with $d\ln N/d\ln a=-3.5$ and a fixed minimum grain size of $1\mu m$. We present the opacity as a function of maximum grain size $a_{\rm max}$ in Fig. \ref{fig:kappa}. In the same figure, we also plot the dust opacity with $d\ln N/d\ln a=-2$ to demonstrate its dependency on the choice of power-law index.

The choice of maximum grain size and the power-law index, $d\ln N/d\ln a$, is crucial for our multiwavelength analysis. In the optically thin regime, the spectral index is primarily determined by the dust opacity slope, $\beta = d\ln \kappa/d\ln \lambda$. 
As is illustrated in Fig. \ref{fig:kappa}, $\beta$ generally decreases with increasing maximum grain size, $a_\mathrm{max}$. In particular, when $a_\mathrm{max}$ exceeds the observing wavelength ($\sim$1 mm), $\beta$ drops sharply from 1.7 to around 0.5 for a standard size distribution slope of $d\ln N/d\ln a = -3.5$ and to nearly zero for a flatter distribution of $d\ln N/d\ln a = -2$.

\section{More on the Bouncing Barrier}
\label{sec:bouncing}

Here, we present a short summary on previous studies of the bouncing physics.

Consider two equal grains of size $a$, each an aggregate of numerous small monomers (of size $a_0$), come into contact with a relative velocity of $v$. The aggregates can be porous, with a volumn-filling factor of $\phi$.

We first consider head-on collisions. 
At impact, a fraction of the initial kinetic energy (and the attractive potential between surfaces) is converted into elastic deformation energy and propagate into the grain body. 
Since the impact time is typically much longer than the sound-crossing time across the body, these elastic waves are dissipated as heat. In addition, some of the kinetic energy is used to compactify the local aggregates. 
Sticking occurs if the velocity is low enough that there is little energy left to pull the grains apart. 
\citet{Chokshi1993} derived a threshold velocity for sticking as (their eq. 29)\footnote{This was claimed to be erroneous by \citet{Dominik1997}. But their expression (eq. 10 in that paper) seems to have the wrong dimension.}
\begin{equation}
    v_{\rm stick,Chokshi} \approx 
    {{4 \gamma^{5/6}}\over{E^{1/3} \, a^{5/6} \, \rho^{1/2}}} \sim 0.2 \cm/\s \, \left({a\over{100\mu m}}\right)^{-5/6}\, 
    \label{eq:chokshi}
\end{equation}
where $\gamma$ is the material surface energy (per area), $E$ its Young's modulus, and $\rho$ its bulk density, and we have evaluated using the values of quartz in \citet{Chokshi1993}. 
Ice particles should stick up to a velocity some 30 times higher, as a result of the larger surface energy and the lower strength (which leads to a larger contact area). 
Collisions with higher velocities should result in inelastic bounce that also act to compactify the grains.

Simulations of compact aggregates support the presence of such a bouncing barrier \citep[e.g.][]{Wada2011}.
Most recently, \citet{Oshiro2025} numerically simulated head-on collisions between compact aggregates and  numerically measured the transition velocity (transition from sticking to bouncing) as
\begin{equation}
    v_{\rm stick,Oshiro} 
    \sim 0.2 \cm/\s \left({a \over{100 \mu m}}\right)^{-9/4} \, \left({\phi\over{0.5}}\right)^{-14}\, .
    \label{eq:oshiro}
\end{equation}
They simulated values of the volumn filling factor $\phi$ within the range [0.4,0.5], with the theoretical maximum value at $\phi = \sqrt{2}\pi/6=0.74$. The steep dependency on porosity arises from the much stronger compressive strength of more densely packed aggregates. These bodies deform less at impact, and  therefore take less energy to pull apart. Such a dependency highlights  the sensitivity of the bouncing barrier on the structural property of the grains. Their expression differs from that in eq. \ref{eq:chokshi} in terms of the size dependence.  But \citet{Oshiro2025} argued that their expression is consistent with some experiments.

The more common type of collision is not head-on, but oblique collision. \citet{Dominik1997} argue that, in this case, motion tangent to the contact surface (rolling, sliding and spinning) dictates the energy loss. 
The dominant motion, they suggest, is the rolling motion, where grains roll over each other -- new contacts form on the leading edge of the deformation, while old contacts breaks at the trailing edge. The two processes are intrinsically asymmetric -- contact breaking requires stretching the contact surface by a critical amount, while contact making does not \citep{Dominik1997}. This  leads to excitation of elastic waves and energy loss into heat. Or, rolling comes with a ``friction''.
When the impact energy is too low, the rolling motion cannot consummate to, say, $90^\circ$, and the grains will stick to each other. 
This criterion sets a maximum velocity for sticking as  \citep{Dominik1997}
\begin{equation}
    v_{\rm stick, Dominik} =
    \sqrt{5\frac{\pi a_{\rm mono} F_{\rm roll}}{m_\mu}}
\approx     0.2 \cm/\s\,  \left(\frac{a}{100\mu m}\right)^{-3/2}\,,
    \label{eq:vbounce}
\end{equation}
where $a_{\rm mono}$ is the size of the monomer, $F_{\rm roll}$ the force needed to roll the monomers, and $m_\mu$ the reduced mass of the two colliders. Parameter values are as those in \citet{Dominik2024}. 
This estimate applies to grains that are largely round and not severely fractal in porosity. 

These scalings may give only an indication of the actual dynamics. One important unknown is grain porosity. While grains may start small and fluffy, continued collisions may well compactify them into dense spheres. Such compact aggregates may resist the making of new contacts and bounce off more easily during collisions.  

Lastly, while laboratory experiments convincingly demonstrate the presence of a bouncing barrier \citep{Blum2008,Guttler2010}, the actual bouncing threshold may depend on a range of parameters (grain porosity, composition, monomer size, etc.). In our work, we will follow \citet{Dominik2024,Guttler2010} in adopting eq. 
\refnew{eq:vbounce} as our bouncing barrier.

\section{Turbulent Stirring}
\label{sec:dust}

Here, we summarize results on grain relative speeds, under the wafting of Kolmogorov turbulence ($V_\alpha$). They have been obtained before \citep{ormel2007}. 
We will be focusing on collisions among equal-sized grains, as they are the most important for grain growth.
For the numerical estimates, we will assume the central star has a mass of $1M_\odot$, and adopt the same disk model as in the main text: $\Sigma_{\rm gas} =\Sigma_0\, r_{\rm au}^{-1}$ with $\Sigma_0 = 2700 \g/\cm^2$ and $r_{\rm au} = (r/{1 {\rm au}})$. We further set $h/r = 0.027 r_{\rm au}^{2/7}$ \citep{Chiang1999}, so that near the disk midplane, $\rho_{\rm gas}  = \Sigma_{\rm gas}/(\sqrt{2\pi}h) = 2.7\times 10^{-9} r_{\rm au}^{-16/7}$. 

Let $L$ be the outer driving  scale of the Kolmogorov turbulence and $\eta$ its inner viscous scale. The Reynolds number at scale $\eta$ is, $Re(\eta) = v_\eta \eta/\nu = Re_{\rm min}$, where $\nu = c_s \lambda_{\rm mfp}$ is the gas kinematic viscosity. Here, $c_s$ the local sound speed,  and $\lambda_{\rm mfp}$ the gas collision mean-free-path. We set $Re_{\rm min} \approx 5000$, as appropriate for the onset of turbulence. The definition of $\alpha$ yields $v_L L = \alpha c_s h$, and we take $L =  \beta\, h$, so $v_L = (\alpha/\beta) c_s$ and $L/v_L = (\beta^2/\alpha)\, P_{\rm orb}$.  The origin of the turbulence is currently unknown, so the values of $\alpha$ and $\beta$ are undetermined. Some previous studies have taken $\beta = \sqrt{\alpha}$ \citep[e.g.][]{Cuzzi2001}. 

In Kolmogorov turbulence, turbulent motion at an intermediate scale $l$ ( where $\eta\leq l\leq L$ ) satisfies $v_l/v_L \sim (l/L)^{1/3}$.  An eddy of size $l$ has a turn-over time (or lifetime) of $t_l = l/v_l$. At the viscous scale, this is 
\begin{equation}
t_\eta = {\eta\over{v_\eta}} = {{\beta^2}\over\alpha} \, P_{\rm orb} \left[{{Re_{\rm min}}\over{Re(L)}}\right]^{1/2}\, ,
    \label{eq:teta}
\end{equation}
and we have $Re(L) = v_L L/\nu = \alpha c_s h/\nu = \alpha h/\lambda_{\rm mfp}$. 
We further estimate the molecular mean-free-path as
\begin{equation}
\lambda_{\rm mfp} = {1\over {n\sigma}} \approx {{2.3 m_H}\over{\rho_{\rm gas}\sigma}} 
\sim 0.7\cm\, \,  r_{\rm au}^{16/7}\, ,
    \label{eq:lambad}
\end{equation} 
where we have adopted a gas-gas collisional cross-section of $\sigma = 2\times 10^{-15}\cm^2$. 

For the gas-grain interaction, we consider the so-called Epstein drag regime, i.e., when the grain size satisfies $a \leq \lambda_{\rm mfp}$. In this case, the coupling time between gas and dust is 
\begin{equation}
    t_{\rm stop} \sim {{\rho_{\rm grain} a}\over{\rho_{\rm gas} c_s}}\, ,
    \label{eq:tstop}
\end{equation}
where $\rho_{\rm grain}$ is the grain's bulk density. We consider only grains in the tightly coupled regime, where the Stokes number $St = \Omega t_{\rm stop}  \approx {{\rho_{\rm grain} a}\over{\Sigma_{\rm gas}}} \ll 1$. 

As grains are wafted by the turbulence, the 
relative velocity between grains of the same size $a$ is primarily set by turbulent eddies that have turn-over time matching the grain's stopping time,   $V_\alpha \sim v_l$ with $l$ defined by $t_l \sim t_{\rm stop}$. Eddies larger than these can strongly entrain grains (so there is little relative motion), and eddies smaller than these fluctuate too fast to stir the grains. We define a special size $a_\eta$, the grain size that is marginally coupled to the smallest eddies ($\eta$),
 \begin{eqnarray}
 a_\eta & = &  {{\rho_{\rm gas} c_s t_\eta}\over{\rho_{\rm grain}}} \nonumber \\ &\sim &  
38\cm\,\times 
r_{\rm au}^{-1/2} \,  \left({{\beta^2}\over\alpha}\right)\, \left({{\alpha}\over{10^{-4}}}\right)^{-1/2} \left({{Re_{\rm min}}\over{5000}}\right)^{1/2} 
 \, .
     \label{eq:aeta}
 \end{eqnarray}
Here, we have taken $\rho_{\rm grain} = 1.7 \g/\cm^3$, $c_s/v_{\rm kep} = h/r$ and $v_{\rm kep} = 30 r_{\rm au}^{-1/2}\, \km/\s$.

Grains with sizes much below $a_\eta$ are largely entrained by even the smallest eddies ($t_{\rm stop} \leq t_\eta$). Their relative velocity can be described as $V_\alpha \approx  v_\eta t_{\rm stop}/t_\eta$.  \citet{Weidenschilling1984} elegantly explained this as resulting from terminal motion in a accelerating/decelerating laminar flow.\footnote{Strictly speaking, this is only non-zero between grains of different sizes. But we ignore this fine point here.}  This is also derived formally by \citet{ormel2007}.

We write the following combined expression for turbulent wafting,
 \begin{eqnarray}
V_\alpha & = & {l\over{t_l}} =  \left( {{\alpha^3}\over{\beta^4}} 
\, St
\right)^{1/2} c_s \propto a^{1/2}\, ,\hskip0.21in
{\rm if\, a \geq a_\eta}
\nonumber \\
& = &  v_\eta {{t_{\rm stop}}\over{t_\eta}} = v_\eta \left({a\over{a_\eta}}\right) \propto a^1 \, ,\hskip0.3in {\rm if\, a \leq a_\eta}
     \label{eq:Valpha12}
 \end{eqnarray}
For the lower range ($a < a_\eta$), the full expression is 
\begin{eqnarray}
V_\alpha & = &  {{\alpha^{3/4} a c_s}\over{\Sigma_{\rm gas}^{3/4}}} \, \left[\rho_{\rm grain} \left({\alpha\over{\beta^2}}\right)^{3/2} \,  \, \left({{\sigma}\over{ 2.3 m_H Re_{\rm min}}}\right)^{1/4}
\right]\nonumber \\
& = & 0.025 \cm/\s \,\times  r_{\rm au}^{15/28} \, \left({a\over{100\mu m}}\right) \nonumber \\
& & \times \left({\alpha\over{10^{-4}}}\right)^{3/4}\,  \left({{\Sigma_0}\over{2700\g/\cm^2}}\right)^{-3/4}\, ,
    \label{eq:valpha3}
\end{eqnarray}
where we have taken $\beta^2 = \alpha$ and $Re_{\rm min}=5000$.

When these grains collide, the outcome depends on how $V_\alpha$ compares against  $v_{\rm stick}$, the velocity threshold below which grains stick. Setting $V_\alpha = v_{\rm stick}$ then yields the limit of grain growth, in presence of the bouncing barrier,
\begin{equation}
    a_{\rm bounce} = {\rm Max} \left({{v_{\rm stick}}\over{v_\eta}} \, a_\eta, a_{\rm b0}\right)\, ,
    \label{eq:abounce0}
\end{equation}
where $a_{\rm b0}$ is defined at which $St (a_{\rm b0}) = (\beta^4/\alpha^2)\, v_{\rm stick}^2/(\alpha c_s^2)$. This differs slightly from that advocated in \citet{Dominik2024}. They write $a_{\rm bounce} = {\rm Max}(a_\eta, a_{\rm b0})$ with $a_{\rm b0}$ defined                to have a Stokes number of $St = v_{\rm stick}^2/(\alpha c_s^2)$. The latter difference disappears if one takes $\beta = \sqrt{\alpha}$. But the former difference remains.

Adopting $v_{\rm stick}$ as in eq. \refnew{eq:vbounce} and $V_\alpha$ as in eq. \refnew{eq:valpha3} (valid for $a < a_\eta$), one obtains the following growth limit due to the bouncing barrier, 
\begin{equation}
a_{\rm bounce,1}  \approx  220 \mu m \, \times r_{\rm au}^{-3/14} \, \left[\left({\alpha\over{10^{-4}}}\right)^{-1}
\left({{\Sigma_0}\over{2700 \g/\cm^2}}\right)\right]^{+3/10}\, .
    \label{eq:abounce}
\end{equation}
In the alternative limit of $a > a_\eta$, it is instead
\begin{equation}
    a_{\rm bounce,2} \approx 25 \mu m \, \times r_{\rm au}^{-1/7} \, \left({\alpha\over{10^{-4}}}\right)^{-1/4}
\left({{\Sigma_0}\over{2700 \g/\cm^2}}\right)^{+1/4}\, .
\end{equation}
The actual value is taken to be the maximum of the two. In either case, the dependency on disk parameters is weak. We find it practical, for most parameters, to adopt $a_{\rm bounce} \approx 100\mu m$.

\end{appendix}
\end{document}